\documentclass[11pt]{article}

\usepackage{amssymb}
\usepackage{amsfonts}
\usepackage{amsmath}
\usepackage{graphicx}
\usepackage{graphicx}
\usepackage{caption}
\usepackage{subcaption}
\usepackage{latexsym}

\usepackage{float}
\newcommand{\bb}{\begin{eqnarray}}
\newcommand{\ee}{\end{eqnarray}}
\newcommand{\beq}{\begin{equation}}
\newcommand{\eeq}{\end{equation}}
\newcommand{\ba}{\begin{array}}
\newcommand{\ea}{\end{array}}

\newtheorem{Remark}{Remark}

\newtheorem{Proposition}{Proposition}

\title{Rational solitons of wave resonant interaction models}

\author{Antonio Degasperis\\
Istituto Nazionale di Fisica Nucleare \\
Dipartimento di Fisica, ``Sapienza" Universit\`a di Roma, Italy\\
E-mail: antonio.degasperis@roma1.infn.it
\and
Sara Lombardo\\
Department of Mathematics and Information Sciences\\
Northumbria University, Newcastle upon Tyne, UK\\
E-mail: sara.lombardo@northumbria.ac.uk}

\date{\today}

\begin{document}

\maketitle

\begin{abstract}
Integrable models of resonant interaction of two or more waves in 1+1 dimensions are known to be of applicative interest in several areas. Here we consider a system of three coupled wave equations which includes as special cases the vector Nonlinear Schr\"{o}dinger equations and the equations describing the resonant interaction of three waves. The Darboux-Dressing construction of soliton solutions is applied under the condition that the solutions have rational, or mixed rational-exponential, dependence on coordinates. Our algebraic construction relies on the use of nilpotent matrices and their Jordan form. We systematically search for \emph{all bounded} rational (mixed rational-exponential) solutions and find, for the first time to our knowledge, a broad family of such solutions of the three wave resonant interaction equations.

\vspace{0.2cm}


\vspace{1cm}

%\noindent \emph{This article is dedicated to the memory of our colleague and friend Sergey Valentinovich Manakov.}
\noindent \emph{Dedicated to the memory of our colleague and friend Sergey Valentinovich Manakov.}
\end{abstract}

\newpage
%%%%%%%%%%%%%%%%SECTION 1%%%%%%%%%%%%%%%%%%%%%%%%

\section{Introduction}
Integrable partial differential equations which model nonlinear wave propagation in 1+1 dimension have been largely investigated because of their applicative relevance. In fact, even if approximate, some of them capture important nonlinear effects. This is because they can be derived, as amplitude modulation equations, by multiscale perturbation methods from various kind of (not necessarily integrable) wave equations with the assumption of weak dispersion and nonlinearity (see for instance \cite {D2009} and references therein). The universality of these integrable models has been well recognized \cite{C1989}, \cite{C1991}. The best known and simplest example of such models is the Nonlinear Schr\"{o}dinger (NLS)  equation for the evolution of the amplitude of a quasi-monochromatic wave with wave number $k$ and frequency $\omega$, as given by the linear dispersion function $\omega=\omega(k)$. Many physical applications require however that integrable models be extended to wave coupling. One important instance regards resonance phenomena. If the dispersion relation allows for resonances, multiscale perturbation methods show that the amplitudes of two or more monochromatic waves couple to each other leading to (possibly integrable) systems of nonlinear partial differential equations. The simplest of such integrable systems is the Vector Nonlinear Schr\"{o}dinger (VNLS)  system of equations, given by the following two coupled equations (a subscript denotes partial differentiation)
\begin{equation}\label{VNLS} 
\begin{array}{l}
u^{(1)}_{t} = i \left[ u^{(1)}_{xx}-2\left( s_{1}\,\left\vert u^{(1)}\right\vert
^{2}+s_{2}\,\left\vert u^{(2)}\right\vert ^{2}\right) u^{(1)}\right]\, \\ [2ex]
u^{(2)}_{t} = i \left[ u^{(2)}_{xx}-2\left( s_{1}\,\left\vert u^{(1)}\right\vert
^{2}+s_{2}\,\left\vert u^{(2)}\right\vert ^{2}\right) u^{(2)}\right] \, 
\end{array}
\end{equation}
where, because of the integrability condition, $s_1^2=s_2^2=1$. This system, also known as Manakov system, follows from the \emph{weak} resonant condition that two quasi-monochromatic waves, with wave-numbers $k_1$ and $k_2$, have the same group velocity, i.e. $\omega'(k_1)=\omega'(k_2)$ ($\omega'(k)=d\omega/dk$). In (\ref{VNLS}) $u^{(1)}(x,t)\,,\,u^{(2)}(x,t)$ are the amplitudes of these two resonant waves. A different type of phenomena occurs when the medium nonlinearity includes quadratic terms and the dispersion relation $\omega(k)$ allows for the two wave numbers $k_1$ and $k_2$ to satisfy the \emph{strong} resonant condition $\omega(k_1+k_2)=\omega(k_1) + \omega(k_2)$. In this case a third wave is generated with amplitude $w(x,t)$ and the three amplitudes $u^{(1)}\,,\,u^{(2)}$ and $w$ couple to each other according to the system of equations
\begin{equation}\label{3W}
\begin{array}{l}
u^{(1)}_{t} =  \left[ -c_1 u^{(1)}_{x}-s_{1}\,w^* u^{(2)} \right ] \;\;\;\; \\[2ex]
u^{(2)}_{t} = \left [-c_2 u^{(2)}_{x}+s_{2}\,w u^{(1)} \right ] \\[2ex]
\; 0 \; \;\; = \; w_{x} + \,s_{1}\,s_{2}\,(c_1-c_2)\,u^{(1) \ast}\,u^{(2)}\,\,.
\end{array}
\end{equation}
In this article we construct particular solutions of both the systems (\ref{VNLS}) and  (\ref{3W}). In the construction method, the physical meaning of the wave amplitudes and of the independent variables $x,t$ does not play any essential role. On the other hand, the results given here are likely to be of applicative relevance in a rather broad range of different physical contexts (f.i. fluid dynamics, nonlinear optics, plasma physics, Bose-Einstein condensate) so it should be kept  in mind that the actual meaning of all variables may vary according to context. In particular, for the system (\ref{3W}), if $x$ is the evolution (f.i. time) variable, then this system  is the the well known 3 wave resonant interaction (3WRI) equation \cite{K76} where the three characteristic velocities are $c_1,c_2,0$; otherwise, if the evolution variable is $t$, this system models the nonlocal interaction of two waves (NL2W) \cite{D2011,BDDR2012}. Here rescaling transformations have been used to give the equations (\ref{VNLS}) and  (\ref{3W}) a neat form in terms of their coefficients.
As for the solutions presented below, we observe that elementary symmetries of equations (\ref{VNLS}) and  (\ref{3W}) (such as gauge transformations and coordinate translations) and linear transformations of the $(x,t)$ plane can be used also to eliminate some of the free parameters which may appear in analytic expressions. Indeed  these parameters will be considered in the following as unessential since they can be easily introduced through simple transformations.
\newline
The kind of solutions we construct here are usually referred to as \emph{rational solitons} with the following specifications: they are solitons since they are spectrally characterized by the vanishing of the continuous spectrum component, however the discrete spectrum eigenvalues are so special that their corresponding solutions have a \emph{rational}  dependence on the variables $x,t$, in contrast with the standard soliton whose expression is given in terms of exponentials. Rational solutions of multicomponent wave equations such as (\ref{VNLS}) and  (\ref{3W}) generically have a dependence on coordinates which is richer than in the scalar case by possibly having a mixed \emph{rational and exponential} expression. Despite this feature, in the following we term \emph{rational solitons} all these kinds of solutions.  Pole singularities in the $x,t$ variables cannot be avoided, these being the zeros of the denominator of the rational expression. However, if these singularities occur only for complex (i.e. strictly non real) values of $x$ and $t$, these solutions are bounded and gain  physical relevance. 
\newline
 Rational solutions of integrable partial differential equations attracted immediate mathematical interest in the 70's, first for the Korteweg-de Vries equation, the motion of the poles being associated with integrable many-body dynamics. Then quite a number of papers have been devoted to rational solutions of various integrable  equations for one dependent variable, say Boussinesq equation \cite{AS1978}, Hirota equation \cite{ASA2010}, Kadomtsev-Petviashvili equation \cite{P1998} and NLS equation (see f.i. \cite{ACA2010}--\cite{G2013}). Recently further investigations of rational solutions were extended to integrable systems of two coupled differential equations. In this direction a number of such solutions have been found for the VNLS (\ref{VNLS}) \cite{GL2011,BDCW2012} and for two coupled Hirota equations \cite{CS2013}. Similar extension has been reported also for three coupled NLS equations \cite{ZL2013}. 
\newline
 The starting motivation of such a surge of research work goes back to the observation by Peregrine \cite {P1983} that the simplest rational solution of the focusing NLS equation may well model an ocean rogue wave (for a recent survey, see \cite{KPS2009}). This solution describes a localized lump over a background with a peak amplitude which is three times higher than the surrounding background itself and with a finite life-time. On the physical side, these new \emph{nonlinear objects} were soon recognized as ubiquitous rather than just ocean events and maritime disasters.  Rogue waves have been  observed not only in water tanks \cite{CHA2011}  but also in fiber optics \cite{KFFMDGAD2010} and in plasma \cite {BSN2011}. They are predicted in the atmosphere \cite{SS2009}, in superfluids \cite {GEKMM2008}, in Bose-Einstein condensates \cite{BKA2009} and in capillary waves \cite{SPX2010}.
\newline
In this paper we systematically search for \emph{all bounded} rational (mixed rational-exponential) solutions of both the VNLS equation (\ref{VNLS}) and, for the first time to our knowledge, of  the 3WRI equation (\ref{3W}). We adopt a formalism such that these two equations are simultaneously treated by using an appropriate Lax pair. Our method of construction is based on the standard Darboux-Dressing transformation (DDT) as presented in \cite{DL2007,DL2009}, and briefly summarized in Section \ref{sec:DDT}. Section \ref{sec:RSOL} describes the algebraic algorithm we use to obtain rational solutions. In Section \ref{sec:SOL} we finally display examples of such solutions. The polynomials which appear in some of the expressions 
%of rational solitons 
are given in Appendix \ref{ap:A}. 

%%%%%%%%%%%%%%%%SECTION 2%%%%%%%%%%%%%%%%%%%%%%%%
\section{Lax pair and Darboux-Dressing transformation} 
\label{sec:DDT}
Equations (\ref{VNLS}) and (\ref{3W}) are integrable models and as such admit a Lax representation (a Lax pair). For convenience, we introduce a Lax pair which combines both models. In subsequent sections we will describe these two dynamics separately.
Let
\begin{equation}\label{lax} 
\psi_x=X\psi\,, \quad\psi_t=T\psi\,, 
\end{equation}
where $\psi$, $X$ and $T$ are $3\times3$ square matrices, $\psi=\psi(x,t,k)$ being a common
solution of the two linear ordinary differential matrix equations
(\ref{lax}) while $X=X(x,t,k)$ and $T=T(x,t,k)$ depend on the variables $x$, $t$ and the complex spectral parameter $k$ according to the definitions
\begin{subequations}
\begin{equation}\label{xLaxoperator}
\begin{array}{l}
X(x,t,k) =ik\sigma+Q(x,t)\,, 
 \end{array} 
\end{equation}
\begin{equation}\label{tLaxoperator}
\begin{array}{l}
T(x,t,k)=\alpha \; T_{nls}(x,t,k)+\beta \; T_{3w}(x,t,k)
 \end{array} 
\end{equation}
\end{subequations}
where $Q(x,t)$ contains the dynamical variables $u^{(1)}(x,t)$ and $u^{(2)}(x,t)$ and introduces  two \emph{signs} $s_1$, $s_2$, $s_1^2=s_2^2=1$
\begin{equation}\label{Q}
Q=\left ( \begin{array} {ccc}
0 & s_1u^{(1)*} & s_2u^{(2)*} \\ u^{(1)} & 0 & 0 \\ u^{(2)} & 0 & 0 \end{array} \right )\;\;
\end{equation}
while $\sigma$ is a constant diagonal matrix defined as
\begin{equation}\label{sigma}
\sigma=\left ( \begin{array} {ccc}
1 & 0 & 0 \\ 0 & -1 & 0 \\ 0 & 0 & -1 \end{array} \right )\,.
\end{equation}
The matrices $ T_{nls}$ and $ T_{3w} $ are defined by
\begin{equation}\label{TNLS}
 T_{nls} = 2ik^2\sigma+2kQ +i\sigma(Q^2-Q_x)
 \end{equation}
  \begin{equation}\label{T3w}
 T_{3w} = 2ikC -\sigma W+\sigma [C,Q]\, ,
 \end{equation}
 where $W$ contains the field $w(x,t)$
 \begin{equation}\label{matrixW}
W=\left ( \begin{array} {ccc}
0 & 0 & 0 \\ 0 & 0 & -s_1w^* \\ 0 & s_2w & 0 \end{array} \right )\,.
\end{equation}
$C$ is a real diagonal matrix
 \begin{equation}\label{matrixC}
C=\left ( \begin{array} {ccc}
0 & 0 & 0 \\ 0 & c_1 & 0 \\ 0 & 0 & c_2 \end{array} \right )\,
\end{equation}
while $\alpha$ and $\beta$ are real parameters such that, for $\alpha=1$, $\beta=0$, (\ref{lax}) is the Lax  pair corresponding to the VNLS (Manakov) equation (\ref{VNLS}), and for $\alpha=0$, $\beta=1$ (\ref{lax}) is the Lax  pair corresponding to the 3WRI equation (\ref{3W}). Indeed the compatibility conditions yield the evolution equations
\begin{equation}\label{3W-NLS}
\begin{array}{l}
u^{(1)}_{t} = i \alpha \left[ u^{(1)}_{xx}-2\left( s_{1}\,\left\vert u^{(1)}\right\vert
^{2}+s_{2}\,\left\vert u^{(2)}\right\vert ^{2}\right) u^{(1)}\right]\,+\, \beta \left[ -c_1 u^{(1)}_{x}-s_{1}\,w^* u^{(2)} \right ] \\[2ex]
u^{(2)}_{t} = i \alpha \left[ u^{(2)}_{xx}-2\left( s_{1}\,\left\vert u^{(1)}\right\vert
^{2}+s_{2}\,\left\vert u^{(2)}\right\vert ^{2}\right) u^{(2)}\right] + \beta \left [-c_2 u^{(2)}_{x}+s_{2}\,w u^{(1)} \right ] \\[2ex]
\; 0 \; \;\; = \;\; \beta \;(w_{x} + \,s_{1}\,s_{2}\,(c_1-c_2)\,u^{(1) \ast}\,u^{(2)})\,\,.
\end{array}
\end{equation}
%Moreover $s_1^2=s_2^2=1$, and $c_1\,,\,c_2$ are real.
\newline
In the search for novel rational solutions of (\ref{3W-NLS}) we use the Darboux-Dressing construction, as developed in \cite{DL2007} (where the interested reader may find additional references). For completeness, we briefly recall here the essential steps towards a new solution, starting from a known (\emph{seed}) solution:
%The Darboux transformation: 
given a solution $u^{(1)}_{0}\,,\,u^{(2)}_{0}\,,\,w_0$ of (\ref{3W-NLS}), let $\Psi_0$ be a corresponding fundamental matrix solution of (\ref{lax}). Then, if $\chi$ is strictly complex ($\chi \neq \chi^*)$,
\begin{equation}\label{DDT}
\Psi(x,t,k)=\left [ \mathbf{1}+\left (\frac{\chi-\chi^\ast}{k-\chi}\right ) P(x,t) \right ]\Psi_0(x,t,k)\,
\end{equation}
is a solution of (\ref{lax}) with
\begin{subequations}\label{new}
\begin{equation}\label{newU}
\left ( \begin{array} {l}
u^{(1)}(x,t) \\ u^{(2)}(x,t) \end{array} \right ) = \left ( \begin{array} {l} u^{(1)}_0(x,t) \\ u^{(2)}_0(x,t) \end{array} \right ) + \frac{2i(\chi-\chi^*) \zeta^*}{|\zeta|^2-s_1|z_1|^{2}-s_2|z_2|^2} \left ( \begin{array} {r}
z_1 \\ z_2 \end{array} \right )  \;\;,
\end{equation}
\begin{equation}\label{newW}
w(x,t)= w_0(x,t) -  \frac{2is_1s_2 (c_1-c_2)(\chi-\chi^*)z_1^* z_2}{|\zeta|^2-s_1|z_1|^{2}-s_2|z_2|^2}\;\;,
\end{equation}
\end{subequations}
where the vector
\begin{equation}\label{zstructure}
 Z(x,t)= \left ( \begin{array} {c}
\zeta (x,t) \\ z_1(x,t) \\ z_2(x,t) \end{array} \right )\,= \Psi_0(x,t,\chi^*) Z_0\;\;
\end{equation}
is a solution of (\ref{lax}) with $k=\chi^*$  ($\text{Im}\chi \neq 0$)
and $Z_0$ is an arbitrary, constant and complex vector. Moreover in (\ref{DDT}) the projector matrix $P(x,t)$ is
\begin{equation}\label{project}
P(x,t)=  \frac{Z Z^{\dagger}}{|\zeta|^2-s_1|z_1|^{2}-s_2|z_2|^2} \left ( \begin{array} {ccc}
1 & 0 & 0 \\ 0 & -s_1 & 0 \\ 0 & 0 & -s_2 \end{array} \right ) \,.
\end{equation}
Here the condition that the parameter $\chi$ is \emph{not real} is crucial. Indeed, the Darboux-Dressing transformation which adds one real pole to the solution $\Psi_0(x,t,k)$ in the $k-$plane at $k=\chi=\chi^*$ is given by a different formula, as detailed in \cite{DL2007}. However this real-pole transformation will not be used here as it yields rational (or semi-rational) solutions which are singular (i.e. unbounded).
The seed solution  $u^{(1)}_{0}\,,\,u^{(2)}_{0}\,,\,w_0$ of (\ref{3W-NLS}) is the plane wave
\begin{subequations}\label{SEED} 
\begin{equation}\label{SEEDU}
\left ( \begin{array} {l}
u^{(1)}_0(x,t) \\[2ex] u^{(2)}_0(x,t) \end{array} \right ) =  \left ( \begin{array} {l}
a_{1} e^{i(q x-\nu_1 t)}\\[2ex] a_{2} e^{-i(q x+\nu_2 t)} \end{array} \right )  \,,
\end{equation} 
\begin{equation}\label{SEEDW}
w_0(x,t) = is_1 s_2 (c_2-c_1) \frac{a_1 a_2}{2q} e^{-i[2q x+ (\nu_2 - \nu_1) t]} \,,
\end{equation}
\end{subequations}
with
\begin{equation}\label{Frequency}
%\left \{ 
\begin{array} {l}
\nu_1 = \alpha [q^2+2( s_1a_1^2 + s_2a_2^2)] + \beta[ c_1 q + s_2 \frac{a_2^2}{2q}(c_1-c_2)] \,,\\[2ex]
 \nu_2 = \alpha [q^2+2( s_1a_1^2 + s_2a_2^2)] + \beta [-c_2 q+ s_1 \frac{a_1^2}{2q}(c_1-c_2)]\, .\end{array} 
 %\right .
\end{equation}
\begin{Remark}
With no loss of generality the amplitudes $a_1$ and $a_2$ can be taken to be real. Moreover, 
%again with no loss of generality because of Galilei invariance, 
because of Galilei invariance, one may choose the wave numbers  $q$ and $-q$ of these two plane-waves, see (\ref{SEEDU}), to have opposite sign.
 \end{Remark}
 
In order to construct the transformation (\ref{new}) in the case where the seed solution $u^{(1)}_{0}$, $u^{(2)}_{0}$, $w_0$ of (\ref{3W-NLS}) is given by (\ref{SEED}), we have to construct first the solution $\Psi_0$ of the Lax equations (\ref{lax}). To this aim we observe that, once (\ref{SEED}) is fixed, the corresponding $Q_0$ and $W_0$ take the form
 \begin{equation}\label{Q0}
Q_0= G\left ( \begin{array} {ccc}
0 & s_1a_1 & s_2 a_2 \\ a_1 & 0 & 0 \\ a_2 & 0 & 0 \end{array} \right )G^{-1}\,,
\end{equation}
\begin{equation}\label{matrixW0}
W_0= G\left ( \begin{array} {ccc}
0 & 0 & 0 \\ 0 & 0 &  i s_2  \frac{a_1 a_2}{2q}(c_2-c_1) \\ 0 &  is_1  \frac{a_1 a_2}{2q}(c_2-c_1) & 0 \end{array} \right ) G^{-1}\,,
\end{equation}
with
\begin{equation}\label{matrixG}
G= \left ( \begin{array} {ccc}
1 & 0 & 0 \\ 0 & e^{i(q x-\nu_1 t)} & 0 \\ 0 & 0 & e^{-i(q x+\nu_2 t)} \end{array} \right )\,.
\end{equation}
It follows then that
\begin{equation}\label{matrixPhi}
\Psi_0(x,t,k)= G(x,t) \Phi(x,t,k)
\end{equation}
and the Lax pair reads
\begin{equation}\label{constantlax} 
\Phi_x=i\Lambda (k) \Phi\,, \quad\Phi_t=-i\Omega (k) \Phi\,, 
\end{equation}
where
\begin{equation}\label{Lambda} 
\Lambda (k)= \left ( \begin{array} {ccc}
k & -i s_1 a_1 &-i s_2 a_2 \\ -i a_1 & -k-q & 0 \\  -i a_2 & 0 & -k+q \end{array} \right )\;\,
\end{equation}
and
\begin{subequations}\label{Omegamatrix} 
\begin{equation}\label{Omega}
\Omega (k)= \alpha  \,\Omega_{nls}(k)  + \beta \, \Omega_{3w }(k)\,,
\end{equation}
with
\begin{equation}\label{Omeganls}
 \Omega_{nls}(k) \!=\! \left ( \begin{array} {ccc}
-2k^2-s_1a_1^2-s_2a_2^2 & is_1 a_1(2k -q)  & is_2 a_2 (2 k +q) \\ ia_1 (2k-q) & 2k^2- q^2-s_1a_1^2 -2 s_2a_2^2 & s_2 a_1a_2   \\ ia_2 (2k +q) & s_1 a_1a_2  &  2k^2 -q^2 -2 s_1a_1^2 - s_2a_2^2  \end{array} \right ),
\end{equation}
\begin{equation}\label{Omega3w}
 \Omega_{3w }(k) \!=\! \left ( \begin{array} {ccc}
0 & -is_1c_1 a_1 & -is_2c_2 a_2 \\ -ic_1a_1  & -c_1(2 k + q ) - s_2 \frac{a_2^2}{2q}(c_1-c_2) & s_2 \frac{a_1a_2}{2q}(c_1-c_2)   \\ -ic_2a_2 & s_1 \frac{a_1a_2}{2q}(c_1-c_2)  & -c_2(2 k - q ) - s_1 \frac{a_1^2}{2q}(c_1-c_2) \end{array} \right )\;\,.
\end{equation}
\end{subequations}
Since 
\begin{equation}\label{commute}
[\Lambda (k)\,,\,\Omega (k)]=0\,,
\end{equation}
the solution $\Psi_0$ has the expression
\begin{equation}\label{PSI0}
\Psi_0(x,t,k)= G(x,t) e^{i(\Lambda (k) x -\Omega (k) t)}\,.
\end{equation}
Finally, the vector $Z(x,t)$, see (\ref{zstructure}), which appears in the Darboux-Dressing transformation (\ref{new}) reads
\begin{equation}\label{zvect}
Z(x,t)= G(x,t) e^{i(\Lambda (\chi^*) x -\Omega (\chi^*) t )}Z_0\,.
\end{equation}
\begin{Remark}
The Darboux-Dressing transformation (\ref{new}) may lead to a singular solution of (\ref{3W-NLS}) due to zeros of the denominator $|\zeta|^2-s_1|z_1|^{2}-s_2|z_2|^2$. The condition that the signs $s_1\,,\,s_2$ are both negative ($s_1=s_2-1$) is sufficient for this solution to be bounded (i.e. nonsingular). Nevertheless we will keep the signs $s_1\,,\,s_2$ arbitrary.
% also in the next section.
\end{Remark}
\begin{Remark}
The parameter $q$, other than the signs $s_1\,,\,s_2$, is expected to be relevant to the stability of the plane wave solution (\ref{SEED}). Despite the importance of this point we do not discuss it here.
 \end{Remark}
 
%%%%%%%%%%%%%%%%%SECTION 3%%%%%%%%%%%%%%%%%%%%%%%%

\section{Rational solutions}
\label{sec:RSOL}
This section outlines the general scheme to construct \emph{all} bounded  solutions of (\ref{3W-NLS}) which are obtained via the Darboux-Dressing method and whose dependence on coordinates is either rational or a mixture of rational and exponential functions. Two subsections are then devoted to systematically compute the explicit expressions of all these solutions.
The starting observation is that no rational dependence on $x\,,\,t$ of the solution (\ref{new}) exists if the two matrices $\Lambda (k)$ and $\Omega (k)$ (for $k=\chi^*$) are similar to a diagonal matrix. Indeed, this statement stems from the expressions (\ref{new}), together with (\ref{SEED}) and (\ref{zvect}), which imply that, in this generic case, the explicit expression  (\ref{new}) of the solution contains only exponential functions of $x$ and $t$. Therefore we find those particular, \emph{critical}, values $k_c$ of $k$, such that the two matrices  $\Lambda (k_c)$ and $\Omega (k_c)$ are instead similar to a Jordan form. Indeed this form is generically the sum of a diagonal matrix and a nonvanishing nilpotent matrix, therefore the starting elementary observation is that, if $N$ is a nilpotent matrix, say $N^{m+1}=0$ and $N^{m}\neq 0$ for an integer $m$, then $\exp(zN)$ is a matrix valued polynomial of $z$ of degree $m$. Moreover, in order to apply the Darboux-Dressing formula (\ref{new}), the critical value $k_c$ is required to be strictly complex, namely to lie off the real axis of the complex $k-$plane. Therefore through our investigation we disregard all those values of $k$ which are real even if the corresponding matrices $\Lambda (k)$ and $\Omega (k)$ are similar to a Jordan form. Though the matrices $\Lambda (k)$ and $\Omega (k)$ play a similar role, it is convenient to focus first on $\Lambda (k)$ and its characteristic polynomial 
\begin{equation}\label{lambdapoly}
P_{\Lambda}(\lambda)= \text{det}[\lambda-\Lambda(k)]= \lambda^3+A_2(k)\lambda^2 +A_1(k)\lambda +A_0(k)
\end{equation}
whose coefficients take the expression (see (\ref{Lambda}))
\begin{align}\label{lambdacoeff}
A_2(k)&=k\,,\quad A_1(k)=-k^2-q^2+s_1a_1^2 +s_2a_2^2\,,\nonumber\\
A_0(k)&=-k^3+k(q^2+s_1a_1^2 +s_2a_2^2)+q(s_2a_2^2-s_1a_1^2 )\,.
\end{align}
%We then use the following proposition:
The following proposition holds true:
\begin{Proposition}
If $\lambda_1(k)\,,\,\lambda_2(k)\,,\,\lambda_3(k)$ are the three roots of the  characteristic polynomial (\ref{lambdapoly}) then a \emph{necessary} condition for $\Lambda (k_c)$  to be similar to a Jordan form $\Lambda_J$,
\begin{equation}\label{lambdacrit}
\Lambda(k_c) = T\,\Lambda_J\,T^{-1}\,,
\end{equation}
is that either one of them, say $\lambda_3$, is simple and $\lambda_1=\lambda_2$ is double, or $\lambda_1=\lambda_2=\lambda_3$. $T$ denotes the similarity transformation matrix. In the first case, $\Lambda(k_c)$ is similar to a Jordan form $\Lambda_J$ if and only if  $\lambda_1\,=\,\lambda_2$ is geometrically simple,
\begin{equation}\label{lambdajordan2}
\Lambda_J = \left ( \begin{array} {ccc}
\lambda_1 &  \mu  & 0 \\ 0 & \lambda_1 &  0 \\ 0 & 0 & \lambda_3 \end{array} \right )\,,\quad\mu \neq 0 \,;
\end{equation}
while in the second case, $\Lambda(k_c)$ is similar to a Jordan form $\Lambda_J$ if $\lambda_1\,=\,\lambda_2\,=\,\lambda_3$ is geometrically simple, 
\begin{equation}\label{lambdajordan3}
\Lambda_J = \left ( \begin{array} {ccc}
\lambda_1 & \mu_1 & 0 \\ 0 & \lambda_1 &  \mu_1 \\ 0 & 0 & \lambda_1 \end{array} \right )\,,\quad\mu _1 \neq 0 \,.
\end{equation}
\end{Proposition}
%Comments: i) 
\begin{Remark}
The case in which $\lambda_1=\lambda_2=\lambda_3$ is geometrically double is the particular case of (\ref{lambdajordan2}) for $\lambda_1=\lambda_3$.
\end{Remark}
\begin{Remark}
%ii)  in this respect we also point out
We point out for future reference that, in our notation (\ref{lambdajordan2} and \ref{lambdajordan3}), for dimensional reason we prefer to leave the entry $\mu$ in (\ref{lambdajordan2}) and $\mu_1$ in (\ref{lambdajordan3}) as free \emph{nonvanishing} parameters rather than giving them the unit value, $\mu=\mu_1=1$,  as commonly in use.
\end{Remark}
As for the second matrix $\Omega (k_c)$, since it commutes with $\Lambda(k_c)$, see (\ref{commute}), it is consequently taken by the same similarity transformation
\begin{equation}\label{omegacrit}
\Omega(k_c) = T\,\widehat{\Omega}\,T^{-1}
\end{equation}
into a matrix $\widehat{\Omega}$ which commutes with $\Lambda_J$ but it is not necessarily a Jordan form. Indeed, if $\omega_1$, $\omega_2$, $\omega_3$ are the three eigenvalues of  $\Omega (k_c)$, in the first case (i.e. $\lambda_1=\lambda_2$) it necessarily follows that $\omega_1\,=\,\omega_2$, so that
 \begin{equation}\label{hatomega2}
\widehat{\Omega} = \left ( \begin{array} {ccc}
\omega_1 & \rho & 0 \\ 0 & \omega_1 &  0 \\ 0 & 0 & \omega_3 \end{array} \right ) \,,
\end{equation}
which is still a Jordan form if $\rho \neq 0$, while in the second case (i.e. $\lambda_1\,=\,\lambda_2\,=\,\lambda_3$)
\begin{equation}\label{hatomega3}
\widehat{\Omega} = \left ( \begin{array} {ccc}
\omega_1 & \rho_1 & \rho_2 \\ 0 & \omega_1 &  \rho_1 \\ 0 & 0 & \omega_1 \end{array} \right ) \,.
\end{equation}
 On the other hand the values of $\rho$ in (\ref{hatomega2}) and of $\rho_1$ and $\rho_2$ in (\ref{hatomega3}) have no a priori conditions.  
\newline
Once a critical value $k_c$ has been found, setting in (\ref{zvect}) $\chi=k_c^*$ yields the expression
\begin{equation}\label{zcrit}
Z(x,t)= G(x,t) V(x,t)\,,\quad V(x,t)= \left (\begin{array}{c} v(x,t) \\
v_1(x,t) \\ v_2(x,t) \end{array} \right ) = T e^{i(\Lambda_J x -\widehat {\Omega} t )}\left (\begin{array}{c} \gamma_1 \\
\gamma_2 \\ \gamma_3 \end{array} \right )  \,,
\end{equation}
where $\gamma_1\,,\,\gamma_2\,,\,\gamma_3$ are arbitrary complex constants.
 Due to the nilpotent part of $\Lambda_J$ and $\widehat{\Omega}$,  this last expression  yields a dependence of $V(x,t) $ on $x$ and $t$ which is partially rational. Indeed, by inserting (\ref{lambdajordan2}) and (\ref{hatomega2}) into (\ref{zcrit}) yields the \emph{semi-rational} dependence 
\begin{equation}\label{zcritrat2}
V(x,t)= T \left ( \begin{array}{c} ( \gamma_1 + \gamma_2 \xi  )  e^{i(\lambda_1 x -\omega_1 t) } \\   \gamma_2  e^{i(\lambda_1 x -\omega_1 t) } \\  \gamma_3 e^{i(\lambda_3 x -\omega_3 t)} \end{array} \right )\,,\quad\xi= i(\mu x - \rho t)
\end{equation} 
in the case $\lambda_3$ and $\omega_3$ are (algebraically) simple. In the alternative case in which $\lambda_1$ and $\omega_1$ are (algebraically) triple, the expression of $V$ follows by using instead (\ref{lambdajordan3}) and (\ref{hatomega3}) and it reads
 \begin{equation}\label{zcritrat3}
V(x,t)= e^{i(\lambda_1 x -\omega_1 t) } T \left ( \begin{array}{c} \gamma_1 + \gamma_2 \xi_1 + \gamma_3 \zeta  \\   \gamma_2 + \gamma_3 \xi_1   \\   \gamma_3 \end{array} \right ) \,,\quad\xi_1=i( \mu_1 x - \rho_1 t) \,,\quad\zeta= \frac12 \xi_1^2  - i\rho_2 t \,.
\end{equation} 
Using (\ref{zcrit}) the expression (\ref{new}) of the solution $u^{(1)}$, $u^{(2)}$, $w$ of (\ref{3W-NLS}) can be written in the more explicit form:
\begin{subequations}\label{NEW}
\begin{equation}\label{NEWU}
 \left ( \begin{array} {l}
u^{(1)}(x,t) \\ u^{(2)}(x,t) \end{array} \right ) = \left ( \begin{array} {cc} e^{i(qx-\nu_1 t)} & 0 \\ 0 & e^{-i(qx+\nu_2 t)} \end{array}\right ) \left [ \left ( \begin{array} {l}a_1 \\ a_2 \end{array} \right ) + \frac{2i(k_c^*-k_c) v^*}{|v|^2-s_1|v_1|^{2}-s_2|v_2|^2} \left ( \begin{array} {r}
v_1 \\ v_2 \end{array} \right ) \right ] 
\end{equation}
\begin{equation}\label{NEWW}
w(x,t)= is_1s_2 (c_2-c_1)e^{-i[2qx+(\nu_2-\nu_1) t]}\left [\frac{a_1a_2}{2q} +  \frac{2(k_c^*-k_c)v_1^* v_2}{|v|^2-s_1|v_1|^{2}-s_2|v_2|^2}\right ]\,.
\end{equation}
\end{subequations}
These last expressions (\ref{NEW}) readily show that, if the three eigenvalues $\lambda_j$ are all the same, $\lambda_1=\lambda_2=\lambda_3$, then the solution (\ref{NEW}) is purely rational as its expression does not contain any exponentials (see (\ref{zcritrat3})). In the alternative case, $\lambda_1=\lambda_2\neq \lambda_3$, the expression (\ref{zcritrat2}) shows that the solution (\ref{NEW}) is generically expressed in terms of both exponential and rational functions.  Non generically, however, the dependence on coordinates is purely rational if $\gamma_3=0$ while it contains only exponentials if $\gamma_2=0$. 
We summarize the step-by-step construction of all such solutions of (\ref{3W-NLS}) as follows: once a critical value  $k_c$ off the real axis is computed, one computes  the corresponding eigenvalues $\lambda_j$, $\omega_j$ and the off-diagonal entries $\rho$ or $\rho_1$, $\rho_2$; the corresponding similarity matrix $T$ is then computed and thus, using the formula (\ref{NEW}), the final expression of the solution. \\
The following two subsections describe the computation of the critical values $k_c$ and of the corresponding similarity transformation matrix $T$.

%%%%%%%%%%%%%%%%SUBSECTION 3.1%%%%%%%%%%%%%%%%%%%%%%%%%%%%%%%%%%%%%%%%%%%
\subsection{The case $\lambda_1=\lambda_2=\lambda_3$}
We start by requiring that the three roots of the characteristic polynomial (\ref{lambdapoly}) 
coincide with each other, namely $P_{\Lambda}(\lambda)=(\lambda-\lambda_1(k))^3$, so that
\begin{equation}\label{3equal}
\lambda_1(k)=\lambda_2(k)=\lambda_3(k)=\text{tr}(\Lambda(k))/3 =-k/3\,.
\end{equation}
 Moreover, by Cayley theorem, $[\Lambda(k)+k/3]^3=0$ (we omit to write the identity matrix $I$ where no confusion can arise). Therefore the requirement that the matrix $[\Lambda(k)+k/3]$ be nilpotent yields the critical values $k_c$. We disregard the case $[\Lambda(k)+k/3]^2=0$ because it leads to the strong reduction $a_1a_2=0$ and to real critical values of $k$. Moreover the condition $[\Lambda(k)+k/3]^2\neq 0$ excludes the limiting case in which (\ref{lambdajordan2}) holds for $\lambda_1=\lambda_3$ (see Remark 4).
This way we compute all critical values $k_c$. By disregarding those values which are real, we are left with one case only, namely 
\begin{equation}\label{3lambda} 
 q\neq 0\,, \quad k_c=i\frac{\sqrt{27}}{2}\epsilon q\,,\quad s_1= s_2=-1\,,\quad a_1=a_2=2 q\,,\quad \epsilon^2=1 \,.
\end{equation}
In this case the critical value $k_c$ is imaginary and the free parameters are $q$ (real) and the sign $\epsilon$; hence the Darboux-Dressing transformation (\ref{new}) applies and the resulting solution will be considered below. 
 \newline
 It now remains to provide the similarity transformation matrix $T$, as well as the two matrices 
$\Lambda_J$ and $\widehat{\Omega}$, namely $\omega_1$ and $\rho_1$, $ \rho_2$ (see (\ref {hatomega3})).
 $T$ is however already given by (\ref{lambdajordan3}) with $\lambda_1(k_c)=-k_c/3$ (the non vanishing parameter $\mu$ may be fixed according to convenience).
Needless to say, the expression of the similarity matrix $T$ is not unique and the one we give below may be changed, for instance, by a multiplication factor.
In the present case in which $\lambda_1=\lambda_2=\lambda_3$ and $\Lambda-\lambda_1$ is nilpotent with $(\Lambda-\lambda_1)^2\neq 0$, $(\Lambda-\lambda_1)^3=0$, the construction of  the similarity transformation matrix $T$ requires a tedious but straight computation and we limit ourselves to give the final formula:
$\lambda_1=\lambda_2=\lambda_3=-i\frac{\sqrt{3}}{2}\epsilon q$ so that 
\begin{equation}\label{Nii} 
\Lambda (k_c)=\lambda_1 + \mu_1 N\,,\quad N= \left ( \begin{array} {ccc}
\epsilon \sqrt{3} & 1 & 1 \\ -1 & \theta & 0 \\ -1 & 0 &  \theta^* \end{array} \right ),\quad 
\mu_1=2iq\,,\quad \theta=\frac12 (-\epsilon \sqrt{3}+i)\,,
\end{equation}
where the dimensionless matrix $N$ is nilpotent and $\theta$ is a phase factor, namely $|\theta|=1$. In this case the similarity transformation (\ref{lambdacrit}), with (\ref{lambdajordan3}), is provided by the  matrix
\begin{equation}\label{Tii} 
T= \left ( \begin{array} {ccc}
  \theta & 0 & -i \\ 1 & \theta^* & i  \epsilon \sqrt{3}  \\ i \theta^*& i  &  0 \end{array} \right )\;\,
\end{equation}
whose \emph{Jordanization} action is specified by the formula
\begin{equation}\label{jordanize}
N=T N_J T^{-1}\,,\quad N_J=  \left(\begin{array}{ccc} 0 &1 & 0 \\ 0 & 0 & 1 \\ 0 & 0 & 0 \end{array} \right )\,.
\end{equation} 
As for the matrix $\Omega$, 
$
\omega_1=\omega_2=\omega_3= \text{tr}(\Omega) /3 =\frac{11}{2} \alpha q^2 +\beta q [c_1-c_2-i\epsilon \sqrt{3}(c_1+c_2)]
$
and 
\begin{eqnarray*} \label{Omegaii} 
&\Omega(k_c)=\omega_1+  2\alpha q^2 \left ( \begin{array} {ccc}
8 & 3\epsilon \sqrt{3}+i & 3\epsilon \sqrt{3}-i \\ -3\epsilon \sqrt{3}-i & -4 &-2 \\ -3\epsilon \sqrt{3}+i & -2 & -4 \end{array} \right )+ \\
 \\ 
 &+  \beta q
 \left ( \begin{array} {ccc}
i\epsilon \sqrt{3} (c_1+c_2)+c_2-c_1 & 2ic_1  & 2ic_2 \\ -2ic_1 & i\epsilon \sqrt{3}(c_2-2c_1)-c_2 &-2(c_1-c_2) \\ -2ic_2 & -2(c_1-c_2) & i\epsilon \sqrt{3}(c_1-2c_2)+c_1 \end{array} \right ) \,, 
\end{eqnarray*}
while $\widehat{\Omega}$ has the expression (\ref{hatomega3}), namely $\widehat{\Omega}=\omega_1 + \rho_1 N_J + \rho_2 N_J^2$ which implies
\begin{equation}\label{Omegaexpr}
\Omega(k_c)= \omega_1 + \rho_1 N + \rho_2 N^2\,,
\end{equation}
where the matrix $N$ has the expression (\ref{Nii}). Comparing (\ref{Omegaexpr})  with (\ref{Nii}) yields 
\begin{equation}\label{rhoii}
\rho_1= 4\alpha q^2 \epsilon \sqrt{3}+2\beta q (\theta c_1-\theta^* c_2)\,,\quad \rho_2= 4\alpha q^2 +2\beta q( c_1-c_2)\;.
\end{equation}
We now apply the Darboux-Dressing construction formula (\ref{NEW}) with the naked solution appropriate to this case (namely (\ref{SEED}) with $a_1=a_2=2q$), and the vector $V(x,t)$ as given by (\ref{zcritrat3}). Thus we arrive at the following expression of the solution:
\begin{subequations}\label{newii}
\begin{equation}\label{newUii}
 \left ( 
\begin{array}{l} u^{(1)}(x,t) \\ u^{(2)}(x,t) \end{array} \right )\! =\! 2q\! \left ( \begin{array} {cc}  e^{i(qx -\nu_1 t)} & 0 \\ 0 & e^{-i(qx+\nu_2 t)} \end{array} \right )\! \!\left [ \!  \left ( \begin{array} {l} 1 \\ 1  \end{array} \right )
 + \frac{3\epsilon \sqrt{3}A^*}{|A|^2 + |A_1|^2 + |A_2|^2}  \left ( \begin{array} {c} \theta^* A_1 \\ \theta A_2 \end{array} \right ) \right ]\!,
\end{equation}
\begin{equation}\label{newWii}
w(x,t)= 2i q(c_2-c_1)e^{-i[2qx + (\nu_2-\nu_1)t]} [1+\frac{3\epsilon \sqrt{3}\theta^* A_1^* A_2}{|A|^2 + |A_1|^2 + |A_2|^2} ]
\end{equation}
\end{subequations}
with the notation
\begin{equation}\label{notationii}
%\left \{ 
\begin{array}{l} 
\nu= -15\alpha q^2 -\frac32 \beta q(c_1-c_2) \,,\quad \nu_1=\nu + \frac12 \beta q (c_1+c_2)\,,\quad \nu_2=\nu - \frac12 \beta q (c_1+c_2)\;,\; \\[2ex]
A=\gamma_1+\gamma_2 \xi_1 + \gamma_3(\zeta-i\theta^*) \;,\\[2ex]
A_1= \gamma_1+\gamma_2 (\xi_1+\theta^*) + \gamma_3(\zeta+\theta^* \xi_1 +i\epsilon \sqrt{3})\,,\quad A_2=\gamma_1+\gamma_2 (\xi_1 + \theta)+\gamma_3(\zeta+\theta \xi_1)\,,  \end{array} 
%\right .
  \end{equation}
while $\xi_1$ and $\zeta$ are defined by (\ref{zcritrat3}) with $\mu_1=2iq$ (see (\ref{Nii})). We observe that not all the three complex parameters $\gamma_1\,,\,\gamma_2\,,\,\gamma_3$, as introduced via (\ref{zcrit}),  are essential as one of them can be arbitrarily fixed and two more real parameters can be absorbed as translations of $x$ and $t$. The analysis of this solution is detailed in section 4.

 %%%%%%%%%%%%%%%%%%%%SUBSECTION 3.2%%%%%%%%%%%%%%%%%%%%%%%%%%%%
 
\subsection{The case $\lambda_1=\lambda_2\neq \lambda_3$}
Here we consider the case in which, for a critical value $k=k_c$, one eigenvalue (say $\lambda_1$) of $\Lambda(k)$ is algebraically double but geometrically simple, so that $\Lambda(k_c)$ is similar to a Jordan form, see (\ref{lambdacrit}) and (\ref{lambdajordan2}). Since finding $k_c$ generically requires computing the roots of a fourth order polynomial (see below), we postpone this computation and we construct first  the similarity transformation matrix $T$ with the assumption that $k=k_c$ is known. If  $\lambda_1=\lambda_1(k_c)$ and $\lambda_3=\lambda_3(k_c)$ are the corresponding eigenvalues of $\Lambda$ we obtain the following general expression of $T$
\begin{equation}\label{T} 
T= \left ( \begin{array} {ccc}
  \phi_1 & \phi_2 & \phi_3 \\[2ex] -\frac{i\phi_1 a_1}{(\lambda_1 + k + q)} &  -\frac{i\phi_2 a_1}{(\lambda_1 + k + q)} +\frac{i\mu \phi_1 a_1}{(\lambda_1 + k + q)^2} & -\frac{i\phi_3 a_1}{(\lambda_3 + k + q)}   \\[2ex]
    -\frac{i\phi_1 a_2}{(\lambda_1 + k - q)}&  -\frac{i\phi_2 a_2}{(\lambda_1 + k - q)} +\frac{i\mu \phi_1 a_2}{(\lambda_1 + k - q)^2}   &  -\frac{i\phi_3 a_2}{(\lambda_3+ k - q)} \end{array} \right )\,,\quad k=k_c\,,
\end{equation}
which turns out to depend on the three complex parameters $\phi_1\,,\,\phi_2\,,\,\phi_3$, arbitrary except for the condition that the matrix $T$ be non singular. Since the determinant 
\begin{equation}\label{detT}
\text{det}\,T=2\phi_1^2 \phi_3q \mu a_1 a_2 \frac{(\lambda_1-\lambda_3)^2 }{[(\lambda_3+k)^2 - q^2] [(\lambda_1+k)^2 - q^2]^2}
\end{equation}
does not dependent on $\phi_2$, we may take $\phi_2=0$ and conveniently set
$\phi_1=(\lambda_1+k)^2 - q^2$ and  $\phi_3=(\lambda_3+k)^2 - q^2$. With this choice of the parameters the matrix $T$ takes the expression
\begin{equation}\label{Tfine} 
T= \left ( \begin{array} {ccc}
 (\lambda_1+k)^2 - q^2 & 0 & (\lambda_3+k)^2 - q^2 \\ -i a_1(\lambda_1 + k - q) & i\mu a_1(\lambda_1 + k - q)/(\lambda_1 + k + q)  &  -i a_1(\lambda_3 + k - q) \\  -i a_2(\lambda_1 + k + q)&  i\mu  a_2(\lambda_1 + k + q)/(\lambda_1 + k - q)  & -i a_2 (\lambda_3 + k + q)  \end{array} \right )\,,\; k=k_c\;,
\end{equation}
where the condition of being invertible reads $q\mu a_1a_2(\lambda_1-\lambda_3)\neq 0.$
We note that the derivation of this expression requires not only that $P_{\Lambda}(\lambda_1)=P_{\Lambda}(\lambda_3)=0$ but also that $P'_{\Lambda}(\lambda_1)=0$ where $P'_{\Lambda}(\lambda)=dP_{\Lambda}(\lambda)/d\lambda$.
Since this matrix $T$ becomes singular (i.e. non invertible) if $q=0$, see (\ref{detT}), before proceeding further we prefer to first consider  this separate case here below.
\newline
The assumption $q=0$ leads to consider two separate cases, namely either $s_1 a_1^2 +s_2 a_2^2 \neq 0$ or $s_1 a_1^2 +s_2 a_2^2=0$. We disregard this second case as our analysis shows that its corresponding solution becomes singular because of the vanishing of the denominator in the expression (\ref{NEW}). Thus we treat here only the case in which $q=0$ and $s_1 a_1^2 +s_2 a_2^2$ is strictly non vanishing.  With these assumptions the explicit expression of the roots of $P_{\Lambda}(\lambda)$ are
\begin{equation}\label{radiciq=0i}
\lambda_1= \sqrt{k^2-s_1 a_1^2 -s_2 a_2^2}\,,\quad \lambda_2= -\sqrt{k^2-s_1 a_1^2 -s_2 a_2^2}\,,\quad \lambda_3=-k\,,\quad q=0\,.
\end{equation}
The conditions that $\lambda_1=\lambda_2$ and that the value of $k_c$ be not real leads to the condition $s_1 a_1^2 +s_2 a_2^2 \,<\,0$.  This therefore excludes the choice $s_1=s_2=1$ and leads to the two values  $k=k_c= ip\,,\,p= \pm \sqrt{-s_1 a_1^2 -s_2 a_2^2}\,,\,\lambda_1=\lambda_2=0\,,\,\lambda_3= - k_c =-ip$. We find however that the condition $s_1 s_2=1$ is necessary and sufficient  for the solution (\ref{newU}) to be non singular (in general singularities come from the zeros of the denominator which appears in this expression). We conclude therefore that only the (focusing) case $s_1=s_2=-1$ is worth considering. Thus in this particular (and interesting, see below) case the eigenvalues are
\begin{equation}\label{q0lambdas}
\lambda_1=\lambda_2=0\,,\quad \lambda_3=-ip\,,\quad p= \epsilon \sqrt{a_1^2 + a_2^2}\,,\quad \epsilon^2=1\;.
\end{equation} 
Thus the matrix $\Lambda$ reads 
\begin{equation}\label{Lambdaq=0} 
\Lambda(k_c)= \left ( \begin{array} {ccc}
ip & ia_1 & ia_2 \\ -ia_1 & -ip & 0 \\ -ia_2 & 0 &  -ip \end{array} \right )
\end{equation}
and is taken into the Jordan form (here we set  $\mu=-ip$, see (\ref{lambdajordan2}))
\begin{equation}\label{LambdaJq=0} 
\Lambda_J=-ip  \left ( \begin{array} {ccc}
0 & 1 & 0 \\ 0 & 0 & 0 \\ 0 & 0 &  1 \end{array} \right )\,, 
\end{equation}
by the similarity transformation (\ref{lambdacrit}) with
\begin{equation}\label{Tq=0} 
T= \left ( \begin{array} {ccc}
  -p & p & 0 \\ a_1 & 0 & a_2   \\  a_2 & 0  &  -a_1 \end{array} \right )\,. 
\end{equation}
Moreover, since this case does not apply to the 3WRI equations (see (\ref{SEEDW})), we set $\alpha=1$ and $\beta=0$ so that the matrix $\Omega(k_c)$ has the expression
\begin{equation}\label{Omegaq=0} 
\Omega(k_c)=  \left ( \begin{array} {ccc}
3p^2 & 2pa_1 & 2pa_2 \\ -2pa_1 & -p^2 +a_2^2 &-a_1a_2 \\ -2pa_2 & -a_1a_2 &  -p^2 +a_1^2  \end{array} \right )\,, 
\end{equation}
which is similar to the Jordan form $\widehat{\Omega}$ (see (\ref{omegacrit}) and (\ref{hatomega2})) with
\begin{equation}\label{omegahatq=0}
\omega_1=\omega_2 = \omega = p^2\,,\quad\omega_3=0\,,\quad\rho= -2 p^2 \,.
\end{equation}
These findings, together with the explicit expression (\ref{zcritrat2}) and the Darboux-Dressing formula (\ref{new}), yield the semi-rational solution of the VNLS equations
\begin{equation}\label{soliton}
 \left ( \begin{array} {l}
u^{(1)}(x,t) \\ u^{(2)}(x,t) \end{array} \right ) = e^{2i\omega t} \left[ \frac{L}{B} \left ( \begin{array} {l}
a_{1} \\ a_{2} \end{array} \right ) + \frac{M}{B} \left ( \begin{array} {r}
a_{2} \\ - a_{1} \end{array} \right ) \right ],
\end{equation}
where
$ L = \frac32 -8\omega^2 t^2 -2 p^2 x^2 +8i\omega t + |f|^2 e^{2p x}$,
$ M= 4f (p x -2i\omega t -\frac 12 ) e^{(p x + i \omega t)}$, 
$ B= \frac12 +8\omega^2 t^2 + 2 p^2 x^2 + |f|^2 e^{2p x}$, and where
$ f $ is a complex arbitrary constant. It should be remarked that the dressing construction   
has introduced $\gamma_1$, $\gamma_2$, $\gamma_3$ as arbitrary parameters, see  
(\ref{zcritrat2}). However only the complex parameter $\gamma_3$ is left essential 
 since the other parameters can be absorbed by  translations of the 
coordinates $x$, $t$. In fact, the expression (\ref{soliton}) is derived by setting $\gamma_1=1/2$, $\gamma_2=1$ and $\gamma_3=-f$. We note also that the dependence 
 of $L, M$ and $B$ (see (\ref {soliton})) on $x,t$ is both polynomial and exponential 
 only through the dimensionless variables $ax$ and $\omega t$. Moreover the vector 
 solution (\ref {soliton}) turns out to be a combination of the two constant orthogonal vectors $(a_1\,,\,a_2)^T$ and $(a_2\,,\,-a_1)^T$.
\newline  
Let us proceed further to the case in which $q\neq0$, and let us maintain the assumption that $k_c$ is known. We first aim to computing the Jordan matrices $\Lambda_J$ (\ref{lambdajordan2}) and $\widehat{\Omega}$ (\ref{hatomega2}), which amounts to computing $\lambda_1\,,\,\lambda_3\,,\,\omega_1\,,\,\omega_3$ and $\rho$. We start from the observation that the eigenvalue $\lambda_1$ is a zero of both the polynomial $P_{\Lambda}(\lambda)$ and of its derivative (see (\ref{lambdapoly}))  $P'_{\Lambda}(\lambda)=3\lambda^2 + 2A_2(k_c) \lambda + A_1(k_c)=3(\lambda-\lambda_+)(\lambda-\lambda_-)$ where 
\begin{equation}\label{lambdaPM}
 \lambda_{\pm}= -\frac13A_2 \pm \sqrt{\left( \frac{A_2}{3}\right)^2 -\frac{A_1}{3}}\;\;.
\end{equation}
Therefore this
readily implies the following proposition: 
\begin{Proposition}
Assume $k=k_c$, then if $P_{\Lambda}(\lambda_+)=0$, the three roots of $P_{\Lambda}(\lambda)$ are 
\begin{equation}\label{rootsP}
 \lambda_1= \lambda_2=\lambda_+\;\;,\;\;\lambda_3= \frac12 (3\lambda_- - \lambda_+)\;\;,
\end{equation}
while if $P_{\Lambda}(\lambda_-)=0$, the three roots of $P_{\Lambda}(\lambda)$ are 
\begin{equation}\label{rootsM}
 \;\;\lambda_1= \lambda_2=\lambda_-\;\;,\;\;\lambda_3= \frac12 (3\lambda_+ - \lambda_-)\;\;.
\end{equation}
\end{Proposition}
The proof of these formulae is elementary and consistent with the fact that the discriminant of a generic third degree polynomial, see (\ref{lambdapoly}), is proportional to the product $[P_{\Lambda}(\lambda_+)][P_{\Lambda}(\lambda_-)]$.
The explicit expression of $ \lambda_1$ and $ \lambda_3$ finally obtains by inserting in (\ref{lambdaPM}) the coefficients $A_2\,,\,A_1$ in terms of $k$ via (\ref{lambdacoeff}).
\newline
 As for  the eigenvalues $\omega_1\,,\,\omega_3$ and of the parameter $\rho$, see (\ref{hatomega2}), we use the similarity property (\ref{omegacrit}), the matrix transformation $T$ being given by  (\ref{Tfine}), and we obtain the expressions (with $k=k_c$)
  \begin{equation}\label{omegaexpr}
% \left \{ 
 \begin{array}{l} 
\begin{array}{l} 
  \omega_1=\omega_2 =  -\alpha\left\{2k \lambda_1+s_1a_1^2+s_2a_2^2+q\left[\frac{s_1a_1^2}{(\lambda_1+k+q)} - \frac{s_2a_2^2}{(\lambda_1+k-q)}\right] \right\}+\\\
\qquad\qquad\;\;-\frac{\beta}{2} \left\{(c_1+c_2)(k-\lambda_1)+(c_1-c_2)\left[\frac{s_1a_1^2 }{(\lambda_1+k+q)}- \frac{s_2a_2^2}{(\lambda_1+k-q)}\right]\right\} \,,
\end{array} 
\\[3ex]
  \begin{array}{ll} \omega_3= & -\alpha\left\{2k \lambda_3+s_1a_1^2+s_2a_2^2+q\left[\frac{s_1a_1^2}{(\lambda_3+k+q)} -\frac{s_2a_2^2}{(\lambda_3+k-q)}\right]\right\} +\\
  &- \frac{\beta}{2} \left\{(c_1+c_2)(k-\lambda_3) +(c_1-c_2)\left[\frac{s_1a_1^2 }{(\lambda_3+k+q)}- \frac{s_2a_2^2 }{(\lambda_3+k-q)}\right]\right\} \end{array} \,,\\[3ex]
  \begin{array}{ll} \rho= & -\alpha \mu \left\{2k-q\left[\frac{s_1a_1^2 }{(\lambda_1+k+q)^2}- \frac{s_2a_2^2 }{(\lambda_1+k-q)^2}\right]\right\} +\\
    & +\frac{\beta}{2} \left\{c_1+c_2 + (c_1-c_2)\left[\frac{s_1a_1^2 }{(\lambda_1+k+q)^2}- \frac{s_2a_2^2 }{(\lambda_1+k-q)^2}\right]\right\} \,.\end{array}
  \end{array} 
  %\right .
 \end{equation}
% \begin{equation}\label{omegaexpr}
% \left \{ \begin{array}{l} 
%  \begin{array}{l} \omega_1=\omega_2 =  -\alpha[2k \lambda_1+s_1a_1^2+s_2a_2^2+q[s_1a_1^2/(\lambda_1+k+q) - s_2a_2^2/(\lambda_1+k-q)] \\
%\quad\quad\quad\quad\;\;     -(\beta/2) \{(c_1+c_2)(k-\lambda_1) +(c_1-c_2)[s_1a_1^2 /(\lambda_1+k+q)- s_2a_2^2 /(\lambda_1+k-q)]\} \end{array} \\[2ex]
%  \begin{array}{ll} \omega_3= & -\alpha[2k \lambda_3+s_1a_1^2+s_2a_2^2+q[s_1a_1^2/(\lambda_3+k+q) -
% s_2a_2^2/(\lambda_3+k-q)] \\
%  & -(\beta/2) \{(c_1+c_2)(k-\lambda_3) +(c_1-c_2)[s_1a_1^2 /(\lambda_3+k+q)- s_2a_2^2 /(\lambda_3+k-q)]\} \end{array} \\[2ex]
%  \begin{array}{ll} \rho= & -\alpha \mu \{2k-q[s_1a_1^2 /(\lambda_1+k+q)^2- s_2a_2^2 /(\lambda_1+k-q)^2]\} \\
%    & +(\beta \mu/2)] \{c_1+c_2 + (c_1-c_2)[s_1a_1^2 /(\lambda_1+k+q)^2- s_2a_2^2 /(\lambda_1+k-q)^2]\} \end{array}
%  \end{array} \right .
% \end{equation}
 The main task now is finding the critical values $k_c$ which are in the complex $k-$plane strictly off the real axis (Im$k_c\neq 0$). These values are zeros of the discriminant  of the polynomial $P_{\Lambda}(\lambda)$ (\ref{lambdapoly}). By taking into account the expression of the coefficients (\ref{lambdacoeff}), this discriminant  turns out to be proportional to the fourth order monodic polynomial
\begin{equation}\label{discri} 
\Delta(k)=k^4 + D_3 k^3 + D_2 k^2 + D_1 k + D_0\,,
\end{equation}
where the coefficients are
\begin{equation}\label{discricoeff}
%\left \{ 
\begin{array}{l} 
D_3=(s_2a_2^2-s_1a_1^2)/(2q) \,,\\[2ex]
D_2=-[8q^4-(s_1a_1^2+s_2a_2^2)^2+20q^2(s_1a_1^2+s_2a_2^2)]/(2^4q^2) \,,\\[2ex]
D_1= -9(s_2a_2^2-s_1a_1^2)(2q^2+ s_1a_1^2+s_2a_2^2)/(2^4 q)\,, \\[2ex] 
D_0 = (q^2 -s_1a_1^2-s_2a_2^2)^3/(2^4q^2) - (\frac34)^3 (s_2a_2^2-s_1a_1^2)^2  \;. \end{array} 
%\right.
\end{equation}
Though the generic fourth degree algebraic equation is solvable, the explicit expression of its solutions is so complicate that its use does not make their computation any easier than just computing them numerically. One exception to this wisdom is the case in which this algebraic equation reduces to a second degree equation. This is the case if we assume the condition $s_1a_1^2=s_2a_2^2$ which implies that $D_1=D_3=0$ with the consequence that the vanishing of the polynomial (\ref{discri}) reads as the second degree equation
\begin{equation}\label{discri2} 
\Delta(k)=R(h)=h^2 + D_2 h + D_0 = 0
\end{equation}
for the new variable  $h=k^2$. Here the coefficients are
\begin{equation}\label{solvcoeff}
%\left \{ 
\begin{array}{l} D_2=-(2q^4-a_1^4+10 s q^2a_1^2)/(2^2q^2)\,, \\[2ex]
 D_0 = (q^2 -2sa_1^2)^3/(2^4q^2)  \;. \end{array} 
 %\right.
\end{equation}
In this special case the reality of $a_1\,,\,a_2$ implies the condition  $s_1=s_2=s$ and $a_1^2=a_2^2$, which has been used to pass from (\ref {discricoeff}) to (\ref{solvcoeff}). The search for the critical values $k_c$ in the parameter space, the parameters being $q\,,\,a_1$ and the sign $s$,  is now simple since the four zeros of the discriminant (\ref{discri}) have the explicit expression
\begin{equation}\label{discrzero}
k=k(\eta_1,\eta_2)= \eta_2\left( -\frac12 D_2+\eta_1\sqrt{\frac14 D_2^2-D_0}\right)^{1/2}\,,\quad \eta_1^2=\eta_2^2=1\,,
\end{equation}
 which is the starting point of our short discussion of the corresponding family of solutions we  present in the subsection 4.2.2. We note here that these  expressions of $k_c$ are explicit because of the assumption $s_1a_1^2=s_2a_2^2\,$. In the generic case in which $q\neq 0$ and $s_1a_1^2-s_2a_2^2\neq 0$, we prefer to compute $k_c$ numerically as roots of the discriminant (\ref{discri}). 
%  \newline
%Once a critical value $k_c$ is known, its corresponding semi-rational solution of the equations (\ref{3W-NLS}) is obtained through the following chain of steps: i) compute $\lambda_1$, $\lambda_2$, $\lambda_3\,$,   as well as $\omega_1$, $\omega_2$, $\omega_3\,$ together with $\rho$, or $\rho_1$, $\rho_2$, as shown in this section 3; ii) compute the matrix $T$ which takes the matrices $\Lambda$ and $\Omega$ in Jordan form; iii) compute the auxiliary vector $V$, see (\ref{zcritrat2}) or (\ref{zcritrat3});
%iv) finally compute the solution $u^{(1)}$, $u^{(2)}$, $w$ via the Darboux-Dressing transformation formula (\ref{NEW}).
% In any case, and to conclude this section, once a critical value $k_c$ is known, its corresponding semi-rational solution of the equations (\ref{3W-NLS}) is obtained through the following chain of steps:
%\begin{itemize}
%\item compute $\lambda_1\,,\,\lambda_2\,,\,\lambda_3\,$,  as well as $\omega_1\,,\,\omega_2\,,\,\omega_3\,$ together with $\rho$, or $\rho_1\,,\,\rho_2$, as shown in this section 3.
%\item compute the matrix $T$ which takes the matrices $\Lambda$ and $\Omega$ in Jordan form.
%\item compute the auxiliary vector $V$, see (\ref{zcritrat2}) or (\ref{zcritrat3}).
%\item compute the solution $u^{(1)}\,,\,u^{(2)}\,,\,w$ via the Darboux dressing transformation formula (\ref{NEW}).
%\end{itemize}

%%%%%%%%%%%%%SECTION 4 %%%%%%%%%%%%%%%%%%%%%%%%%%%%%%%%%%%%%%%%%%%%%%%%%%%%%%%%%%%%%%%%%%%%%%%%%%%%%%%%%%%

\section{Analysis of the solutions and conclusions}
\label{sec:SOL}
 In the previous section we have shown the way of deriving a rich family of solutions of the system (\ref{3W-NLS}). In fact we have constructed \emph{all} the bounded (rational or semi-rational) solutions which can be obtained via the DDT method. The aim of this section is to select and detail  some of such solutions. We separately treat those which are solutions of the VNLS system (\ref{VNLS}) (by setting $\alpha=1$, $\beta=0$) and those which are solutions of the 3WRI equations (\ref{3W}) (by setting $\alpha=0\,,\,\beta=1$).
 As for the parameters which appear in the expressions of our solutions, some of them are structural coefficients which enter the partial differential equations (\ref{3W-NLS}), say the signs $s_1$, $s_2$ and the characteristic velocities $c_1\,,\,c_2$; other parameters, i.e. $q$, $a_1$, $a_2$, originate from the background (see (\ref {SEED})) while others, $\gamma_1$, $\gamma_2$, $\gamma_3$, come from the DDT transformation. In this transformation there appears also the critical value $k_c$ of the spectral variable $k$, which depends only on $s_1$, $s_2$, $q$, $a_1$, $a_2$. Although some of the parameters are not essential as they could be eliminated by using simple symmetries, in some cases we prefer to keep them because of their physical significance. We point out also that the background parameter $q$ plays a distinctive role in our solutions as it has no counterpart in the scalar NLS equation.
 
 %%%%%%%%%%%%SUBSECTION 4.1 %%%%%%%%%%%%%%
 
 \subsection{ $\lambda_1=\lambda_2= \lambda_3$}
 \label{triple}
 
 In this case the solutions are rather peculiar as they are all purely rational. Only two critical values of $k$ are possible, namely $k_c=\pm i q\sqrt{27}/2 $ as specified by (\ref{3lambda}). These solutions exist only if $s_1=s_2=-1$, which is the focusing case of the VNLS equations, together with the condition $a_1=a_2=2q$ for the background amplitudes. The general expression of the corresponding solutions is (\ref{newii}).  As for the three complex parameters $\gamma_1$, $\gamma_2$, $\gamma_3$, we omit considering $\gamma_2=\gamma_3=0$ since in this case the expression (\ref{newii}) is trivially that of a plane wave. Thus  we find it  convenient to illustrate the dependence of the solution on these parameters by considering separately the two cases: i) $\gamma_3=0$ and ii) $\gamma_2=0$. With no loss of generality because of translation invariance, one can set $\gamma_2=1\,,\,\gamma_1=0$  in the first case and $\gamma_3=1\,,\,\gamma_2=0$, while $\gamma_1$ remains arbitrary and complex, in the second case. Moreover the expression of the solution is the ratio of two polynomials of second degree in the first case i), and of two polynomials of fourth degree in the second case ii). Figures \ref{figure1} to  \ref{figure4} illustrate these two cases separately for the VNLS and for the 3WRI equations.

%%%%%%%%%%%%%%%%%%%%%%%% VNLS %%%%%%%%%%%%%%%%%%%%%%%%%
 \subsubsection{Solutions of the VNLS} 
Let $X=qx$ and $T=q^2t $ be rescaled variables; let $u^{(j)}(x,t)=q U^{(j)}(X,T)$, $j=1,2$.\\

\begin{figure}[h!]
 \begin{centering} 
 \includegraphics[scale=0.8]{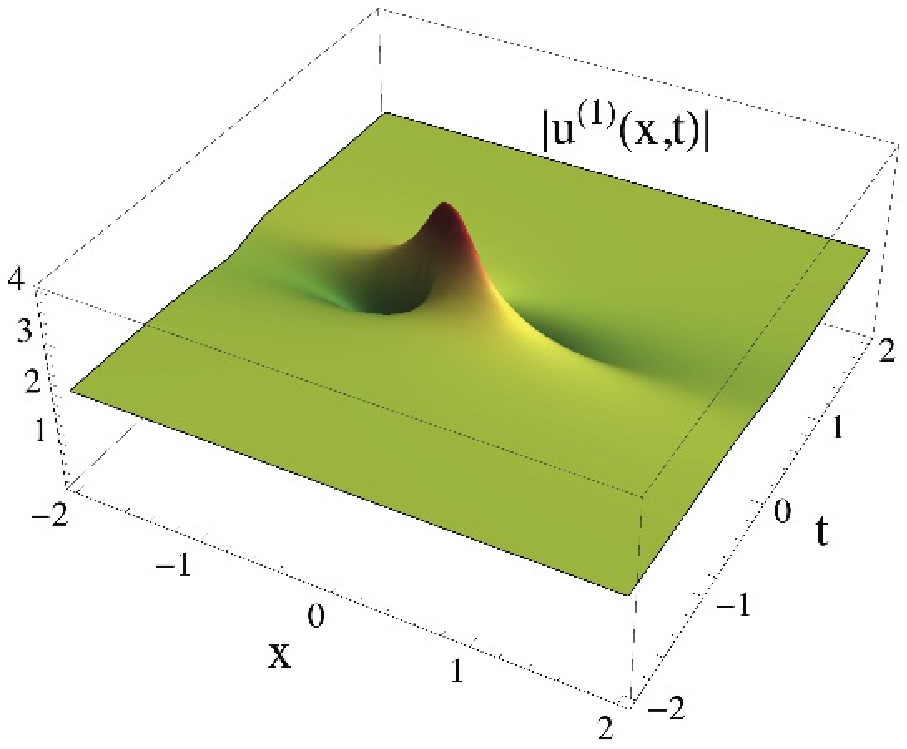} 
%\hspace{0.09cm}
 \includegraphics[scale=0.8]{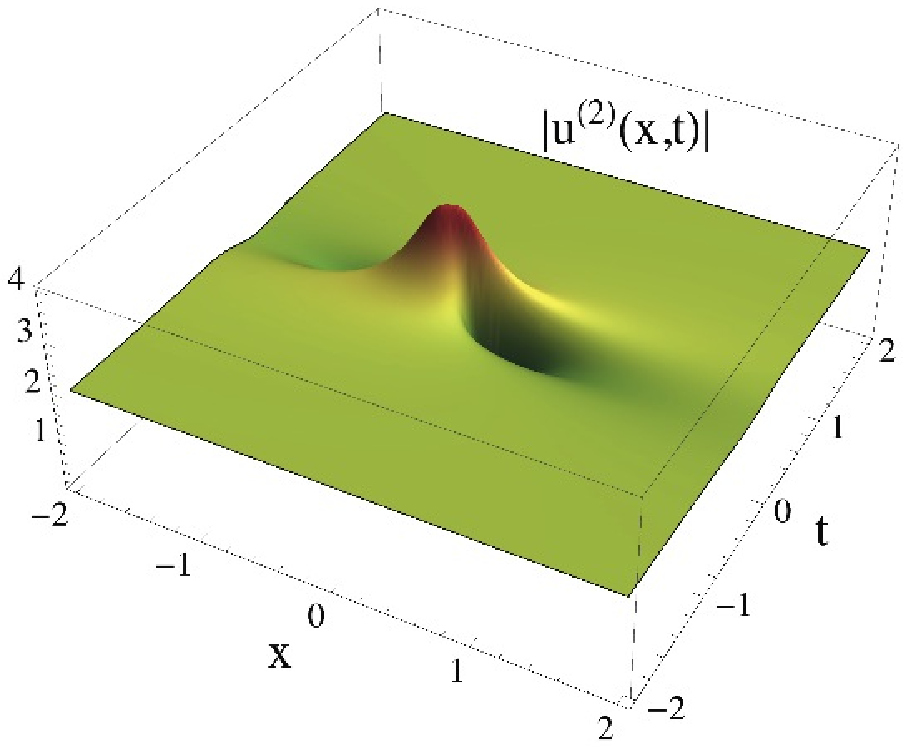}
 \caption{VNLS: $k_c=i\frac{\sqrt{27}}{2}$, $\lambda_1=\lambda_2=\lambda_3\,, s_1= s_2=-1$, $a_1=a_2=2 $, $q=1$, $\epsilon=1$;  $\gamma_2 =1$, $\gamma_1 =\gamma_3 =0$.}
 \label{figure1}
 \end{centering} 
 \end{figure}

\noindent Case $\gamma_3=0\,,\,\gamma_2=1\,,\,\gamma_1=0$
\begin{equation}\label{gamma2VNLS}
U^{(1)}\!=\!2i\theta e^{i(X+15T)} \!\left [\frac{12X^2+ 144 T^2+(4 \epsilon \sqrt{3} +6 i) X -36 i T -1+i \epsilon \sqrt{3}}{12X^2+ 144 T^2 +4 \epsilon \sqrt{3} X +2}\right ].
\end{equation}
Since this solution satisfies the relation $u^{(2)}(x,t)=u^{(1)*}(x,-t)$ we report only the component $u^{(1)}(x,t) =q U^{(1)}(qt,q^2 t)$;  figure \ref{figure1} displays the amplitudes $|u^{(1)}(x,t)|$ and $|u^{(2)}(x,t)|$ for a choice of parameters (see caption). 

\noindent Case $\gamma_3=1\,,\,\gamma_2=0\,,\,\gamma_1\neq 0$
\begin{equation}\label{gamma3VNLS}
U^{(1)}\!=\!2i\theta e^{i(X+15T)} \frac{P^{(1)}_4 }{P_4}\,,\quad
U^{(2)}\!=\!-2i\theta^{\ast} e^{-i(X-15T)} \frac{P^{(2)}_4 }{P_4}
\end{equation}
where the fourth degree polynomials ${P^{(1)}_4}\,,\,{P^{(2)}_4}\,,\,P_4$ are given in Appendix A.
Figure  \ref{figure2} displays the amplitudes $|u^{(1)}(x,t)|$  and $|u^{(2)}(x,t)|$ (see caption). 

%%%%%%%%%%%%%%%%figure 1%%%%%%%%%%%%%%%%%%%

%%%%%%%%%%%%% figure 2 %%%%%%%%%%%%%%%%%%%%%%%
\begin{figure}[h!]
 \begin{centering} 
 \includegraphics[scale=0.8]{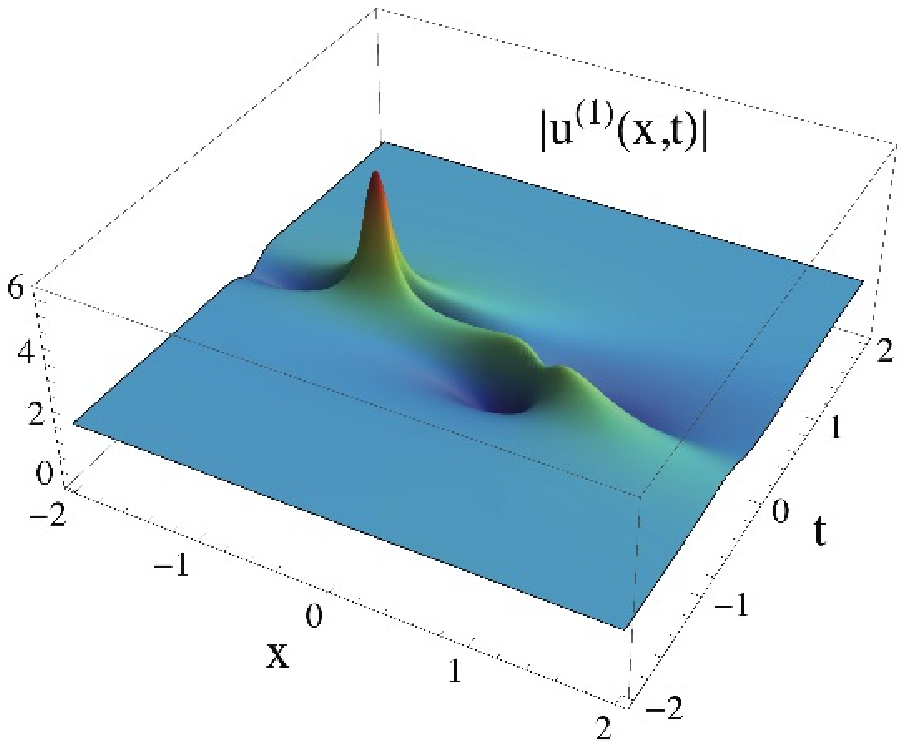} 
%\hspace{0.09cm}
 \includegraphics[scale=0.8]{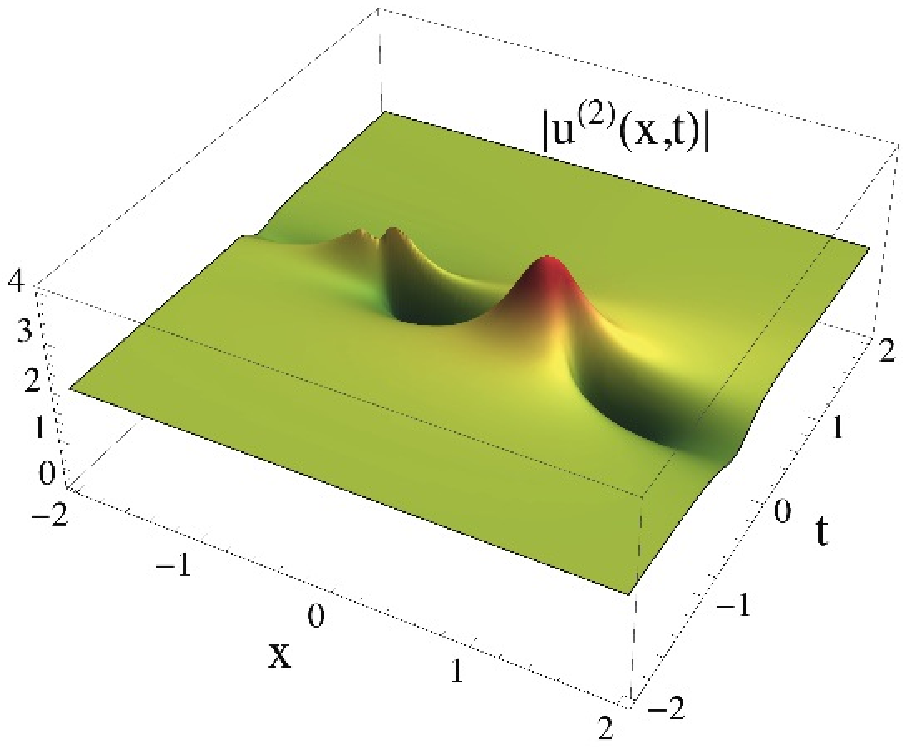}
 \caption{VNLS: $k_c=i\frac{\sqrt{27}}{2}$, $\lambda_1=\lambda_2=\lambda_3\,,s_1= s_2=-1$, $a_1=a_2=2 $, $q=1$, $\epsilon=1$;  $\gamma_1 =i$, $\gamma_2 =0$, $\gamma_3 =1$.}
 \label{figure2}
 \end{centering} 
 \end{figure} 

%%%%%%%%%%%%%%%%%%%%%%% 3WRI %%%%%%%%%%%%%%%
 \subsubsection{Solutions of the 3WRI}
Let $X=qx$ and $T=qt $ be rescaled variables; let 
$u^{(j)}(x,t)=q U^{(j)}(X,T)$, $j=1,2$, $w(x,t)=qW(X,T)$.\\

\begin{figure}[H]
 \begin{centering} 
 \includegraphics[scale=0.53]{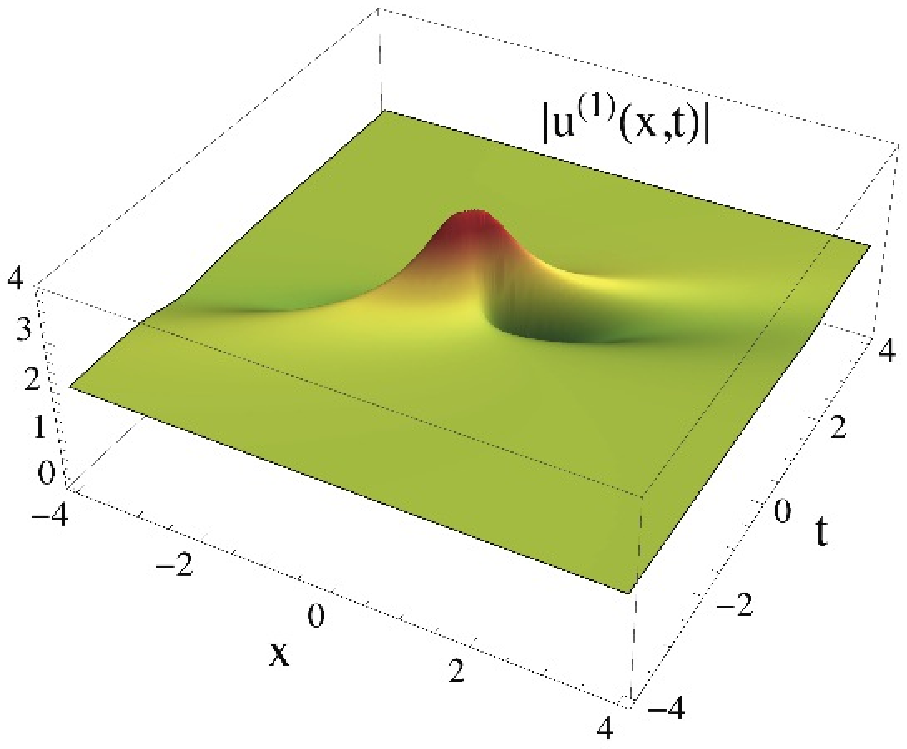} 
%\hspace{0.09cm}
 \includegraphics[scale=0.53]{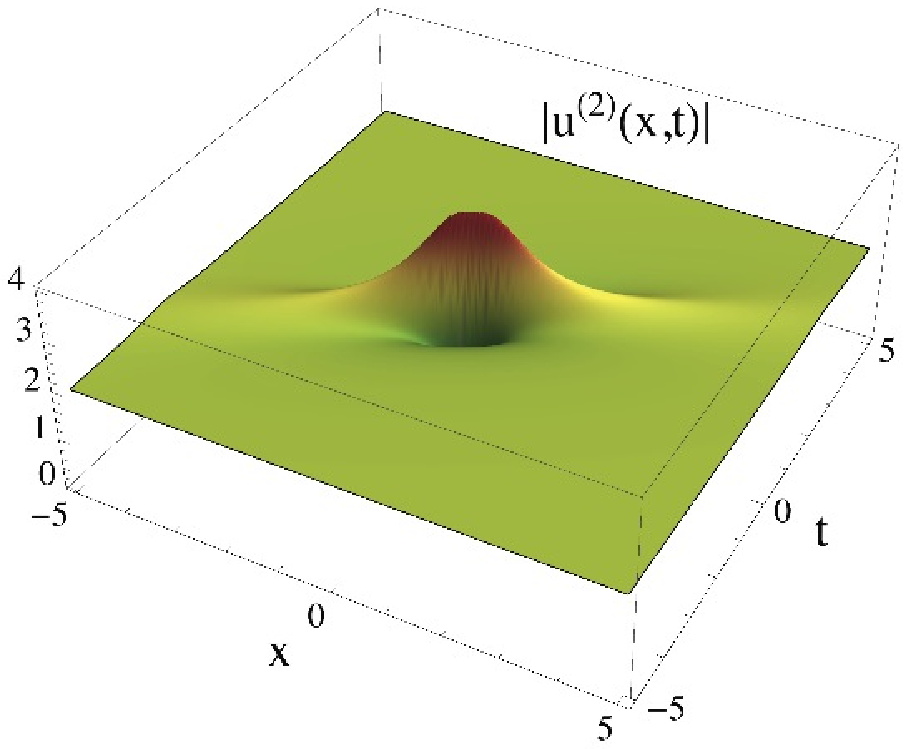}
% \hspace{0.09cm}
 \includegraphics[scale=0.53]{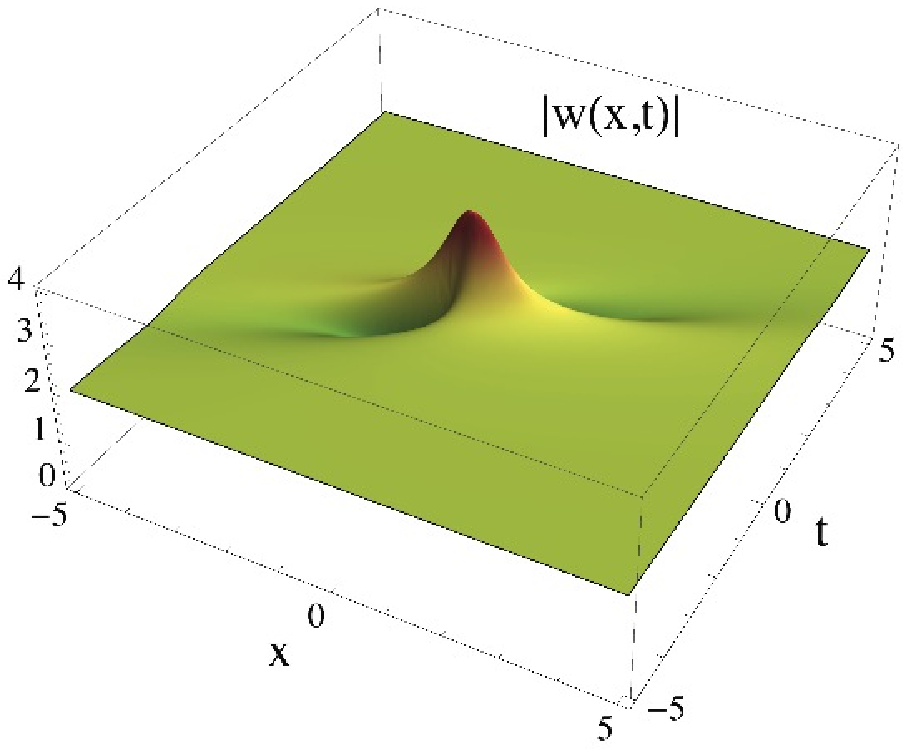}
 \caption{3WRI: $k_c=i\frac{\sqrt{27}}{2}$, $\lambda_1=\lambda_2=\lambda_3\,,s_1= s_2=-1$, $a_1=a_2=2 $, $q=1$, $\epsilon=1$;  $\gamma_2 =1$, $\gamma_1 =\gamma_3 =0$.}
 %; max value .} 
 \label{figure3}
 \end{centering} 
 \end{figure}

\noindent Case $\gamma_3=0\,,\,\gamma_2=1\,,\,\gamma_1=0$
\begin{equation}\label{gamma23W}
U^{(1)}=2i\theta e^{i[X+T(c_1-2c_2)]} \frac{Q^{(1)}_2}{M_2}\,,\quad
W=2\theta (c_1-c_2)e^{-i[2X-T(c_1+c_2)]}\frac{Q_2}{M_2}\,,
%W=2i (c_2-c_1)e^{-i[2X-T(c_1+c_2)]}\frac{Q_2}{M_2}
\end{equation}
where the second degree polynomials $Q^{(1)}_2$, $Q_2$, $M_2$ are given in Appendix A. 
Since this solution satisfies the relation $u^{(2)}(x,t,c_1,c_2)=u^{(1)*}(x,t,c_2,c_1)$ we report the expression of the components $u^{(1)}(x,t) =q U^{(1)}(qx,qt)$, $w(x,t)=qW(qx,qt)$ only. 
Figure \ref{figure3} displays the amplitudes $|u^{(1)}(x,t)|$, $|u^{(2)}(x,t)|$ and $|w(x,t)|$ (see caption).

\noindent Case $\gamma_3=1\,,\,\gamma_2=0\,,\,\gamma_1\neq 0$
\begin{align}\label{gamma33W}
&U^{(1)}=2i\theta e^{i[X+T(c_1-2c_2)]} \frac{Q^{(1)}_4}{M_4}\,,\quad
U^{(2)}=-2i\theta^* e^{-i[X+T(c_2-2c_1)]} \frac{Q^{(2)}_4}{M_4}\,,\nonumber\\  
&W=2\theta (c_1-c_2)e^{-i[2X-T(c_1+c_2)]}\frac{Q_4}{M_4}\,,
\end{align}
where the fourth degree polynomials $Q^{(1)}_{4}$, $Q^{(2)}_{4}$, $Q_{4}$, $M_4$ are given in Appendix A. Figure  \ref{figure4} displays the amplitudes $|u^{(1)}(x,t)|$, $|u^{(2)}(x,t)|$ and $|w(x,t)|$ (see caption).
 
%%%%%%%%%%%%%%%%%%%% figure 3 %%%%%%%%%%%%%%%

 %%%%%%%%%%%%%%%%%%%% figure 4 %%%%%%%%%%%%%%%%%
\begin{figure}[H]
 \begin{centering} 
 \includegraphics[scale=0.53]{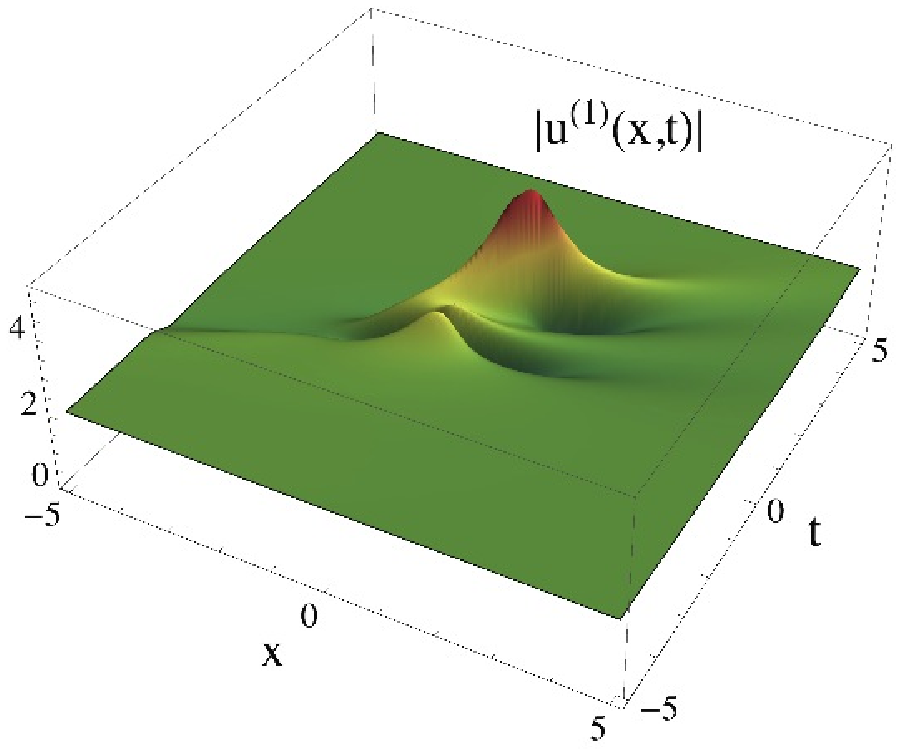} 
%\hspace{0.09cm}
 \includegraphics[scale=0.53]{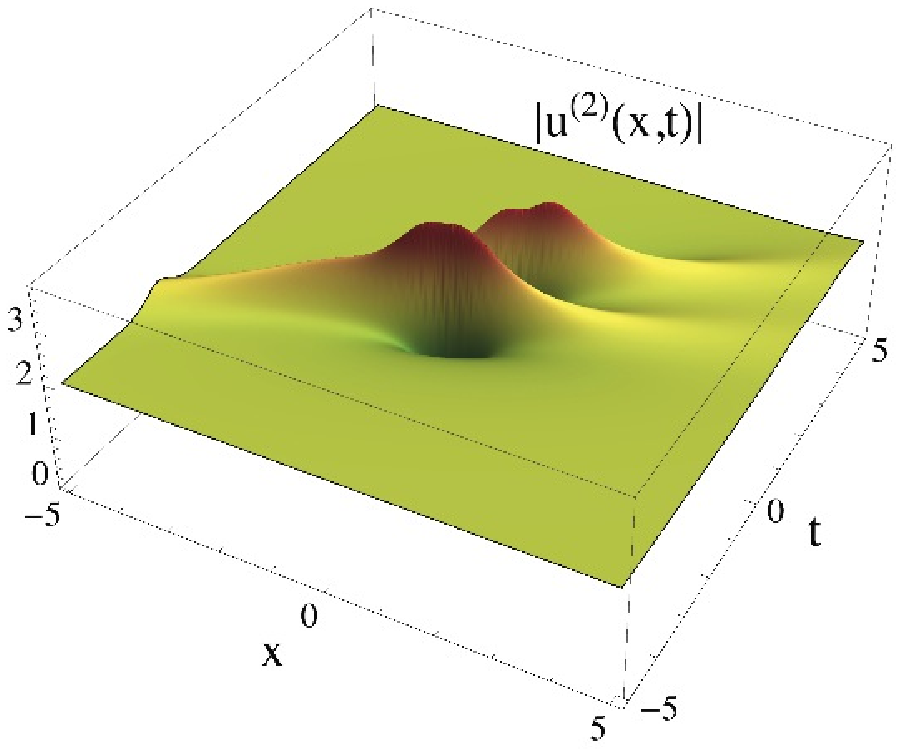}
% \hspace{0.09cm}
 \includegraphics[scale=0.53]{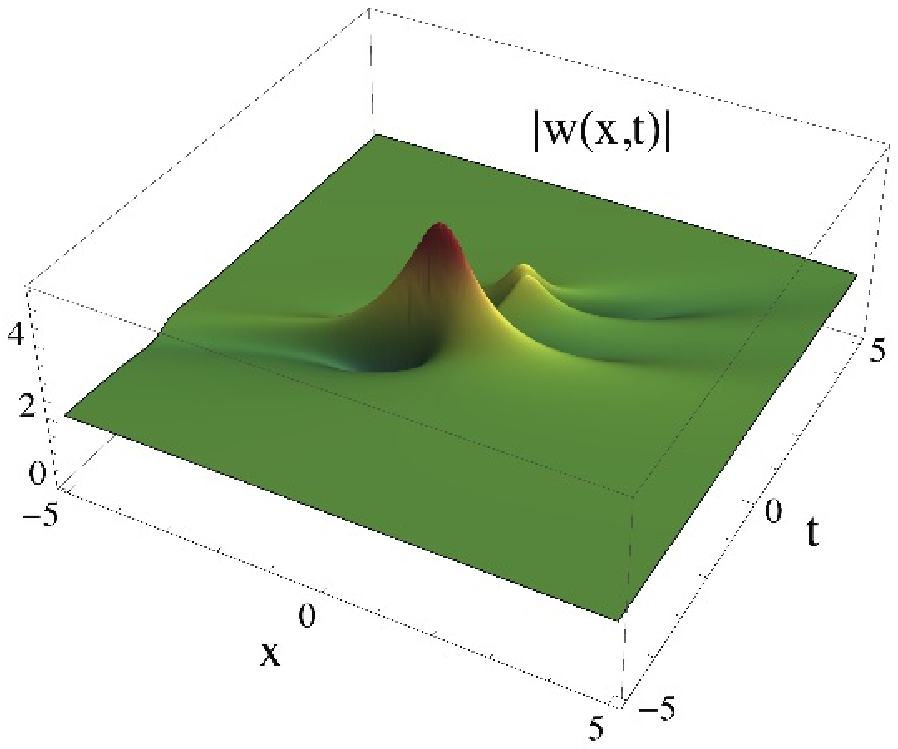}
 \caption{3WRI: $k_c=i\frac{\sqrt{27}}{2}$, $\lambda_1=\lambda_2=\lambda_3\,,s_1= s_2=-1$, $a_1=a_2=2 $, $q=1$, $\epsilon=1$;  $\gamma_1 =i$, $\gamma_2 =0$, $\gamma_3 =1$.}
 %; max value .} 
 \label{figure4}
 \end{centering} 
 \end{figure}

%%%%%%%%%%%%%%SUBSECTION 4.2%%%%%%%%%%% 

 \subsection{The case $\lambda_1=\lambda_2\neq \lambda_3$}
 \label{double}
 The  expression (\ref{NEW}), together with (\ref{zcritrat2}),  shows that generically
 these solutions feature a  dependence on coordinates which is both rational and exponential. In particular, however, if  $\gamma_3 = 0$ the dependence is purely rational while if $\gamma_2 =0$ the solution has only exponential functions. In the following we disregard this last case and consider only solutions with $\gamma_2\neq 0$.  Here we separately consider solutions corresponding to $q=0$ and different background amplitudes, $a_1\neq a_2$,  with $q\neq 0$ but $a_1=a_2$ and, finally, with $q\neq 0$ and $a_1\neq a_2$. These distinctions are merely due to computational reasons. However, and interestingly enough, we numerically show below that in the last two cases (i.e. with $q\neq 0$)  bounded rational solutions exist not only in the focusing case $s_1=s_2=-1$, as for the Peregrine soliton of the scalar NLS equation, but also in the defocusing case $s_1=s_2=1$ and in the mixed case $s_1s_2=-1$. 

%%%%%%%%%%%%%%SUBSECTION 4.2.1%%%%%%%%%%%%
 
 \subsubsection{q=0 and vector Peregrine solutions}
\begin{figure}[H]
\begin{centering} 
 \includegraphics[scale=0.8]{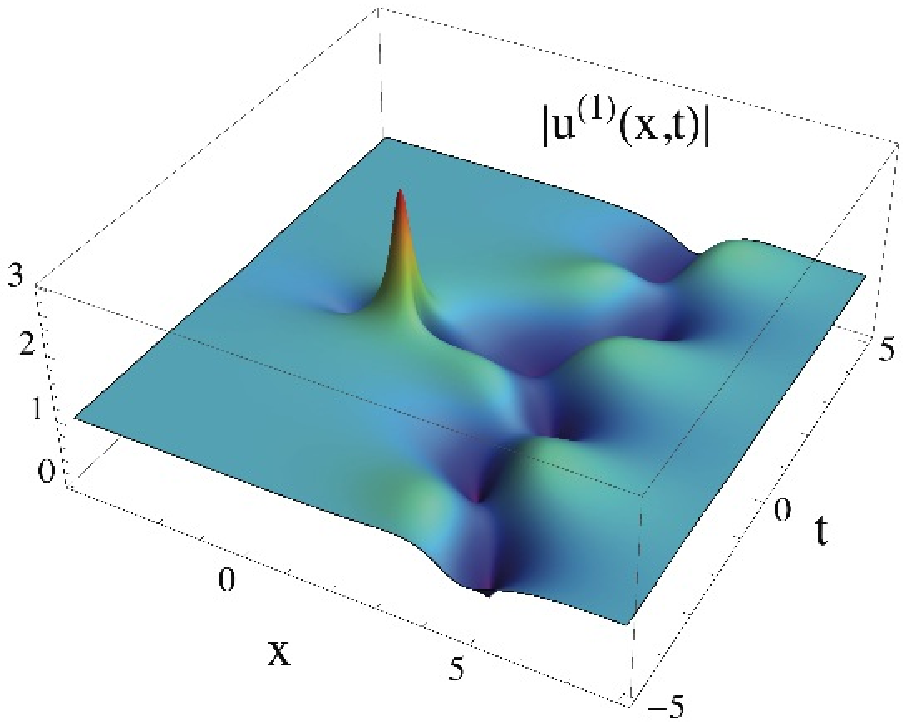} 
%\hspace{0.09cm}
 \includegraphics[scale=0.8]{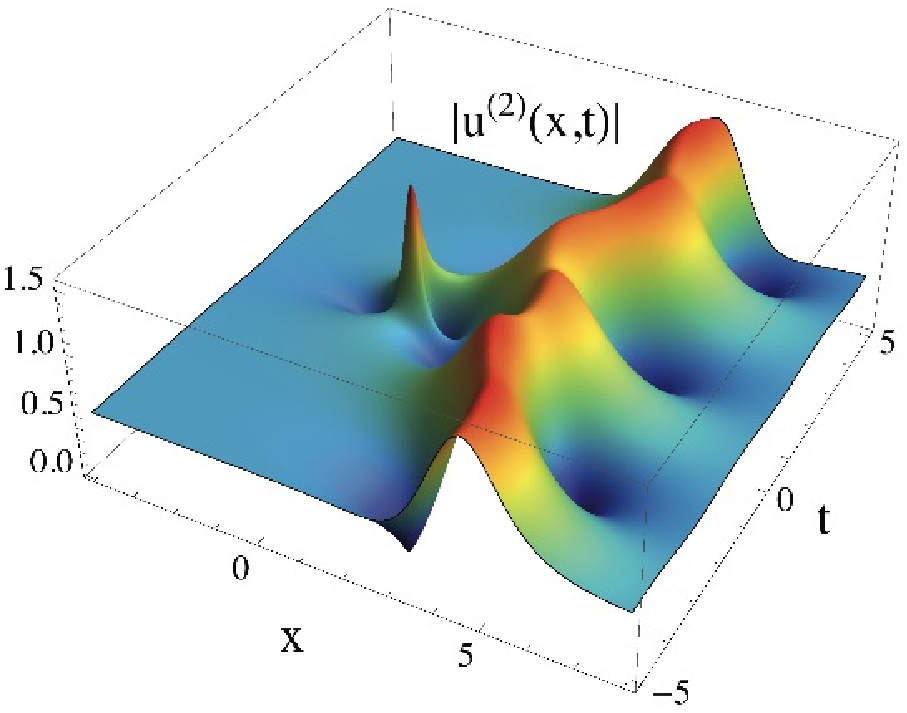}
 \caption{VNLS: $k_c=i\frac{\sqrt{5}}{2}$, $\lambda_1=\lambda_2\neq\lambda_3$, $s_1= s_2=-1$, $q=0, a_1=1$, $a_2=0.5 $, $f=0.1 i$.}
  \label{figure5}
 \end{centering} 
 \end{figure}

 In this case the solution, which is well described by its expression (\ref{soliton}), applies only to the VNLS equation. In this respect we first notice that this expression  (\ref{soliton}), with $f=0\;,$ and $\; a_2=0$, coincides with the Peregrine soliton of the scalar NLS equation. We further note that, since the two components $u^{(1)}(x,t,a_1,a_2)\;,\;u^{(2)}(x,t,a_1,a_2)$  are related to each other by the relation $u^{(2)}(x,t,a_1,a_2)=u^{(1)}(x,t,a_2,-a_1)$, we limit our attention only to $u^{(1)}(x,t)$. In the rescaled variables $u^{(1)}(x,t)=U^{(1)}(X,T)$, $X=x\sqrt{a_1^2+a_2^2}$, $T=t(a_1^2+a_2^2) $, this solution (see (\ref{soliton})) may be written as
 \begin{equation}\label{q0VNLS}
 \begin{array}{l}
U^{(1)}=e^{2iT} \left \{ a_1\left [\frac{(2+8iT)+(4X^2+16T^2-8iT-1)\tanh(X-Z)}{4X^2+16T^2+1} \right ] +a_2\frac{\sqrt{2}f}{4|f|}\left(\frac{8X-16iT-1}{\sqrt{4X^2+16T^2+1}}\right) \frac{1}{\cosh(X-Z)}\right \} \end{array}
\end{equation}
%\begin{equation}\label{q0VNLS}
%\begin{array}{lll}U^{(1)}&=&e^{2iT} \left \{ a_1\left [\frac{(2+8iT)+(4X^2+16T^2-8iT-1)\tanh(X-Z)}{4X^2+16T^2+1} \right ] +\right.\\
% & + &\left. a_2\frac{\sqrt{2}f}{4|f|}\left(\frac{8X-16iT-1}{\sqrt{4X^2+16T^2+1}}\right) \frac{1}{\cosh(X-Z)}\right \} \end{array}
%\end{equation}
where the curve $X=Z(T)$ is the \emph{trajectory} of the soliton as implicitly defined by the formula
\begin{equation}\label{traj}
2|f|^2 e^{2Z}=4Z^2+16T^2+1\,.
\end{equation}
As a consequence of these expressions, the large $T$ asymptotic behavior along the curve $X=Z(T)$ is found to be
\begin{subequations}\label{asym}
\begin{equation}\label{Uasym}
 U^{(1)}(X,T)\rightarrow e^{2iT}\left [a_1\tanh(X-Z)-i a_2 \frac{\sqrt{2}f}{|f|} \text{sign}T \frac{1}{\cosh(X-Z)}\right ]\,,\quad T\rightarrow \pm \infty \,,
 \end{equation}
 \begin{equation}\label{Zasym}
 Z(T)\rightarrow \log |T| + \frac12 \log\left(\frac{8}{|f|^2}\right) + \text{O}\left (
\frac{\log|T|}{|T|} \right),\quad T\rightarrow \pm \infty \,.
\end{equation}
\end{subequations}
 We observe that, as suggested by (\ref{q0VNLS}) and explicitly indicated by the asymptotic expression (\ref{Uasym}), the amplitude $a_1$ multiplies a kink-type profile while the amplitude $a_2$ multiplies a bright-type pulse. Moreover the asymptotic motion (\ref{Zasym}) is that of a particle which comes from $x=+\infty$ and goes back to $x=+\infty$ where it ``stops" since its velocity asymptotically vanishes, namely $dZ(T)/dT\rightarrow 1/T +  \text{O} (
\log|T| / T^2)$.  Figure \ref{figure5} shows an instance (see caption) of the amplitudes $|u^{(1)}(x,t)|$  and $|u^{(2)}(x,t)|$. 
Further instances of this solution (\ref{q0VNLS}) are reported in \cite{GL2011,BDCW2012}.

%%%%%%%%%%%%%%SUBSECTION 4.2.2%%%%%%%%%%%%%

 \subsubsection{$q\neq 0$ and $a_1=a_2$}
 
\begin{figure}[H]
 \begin{centering} 
 \includegraphics[scale=0.8]{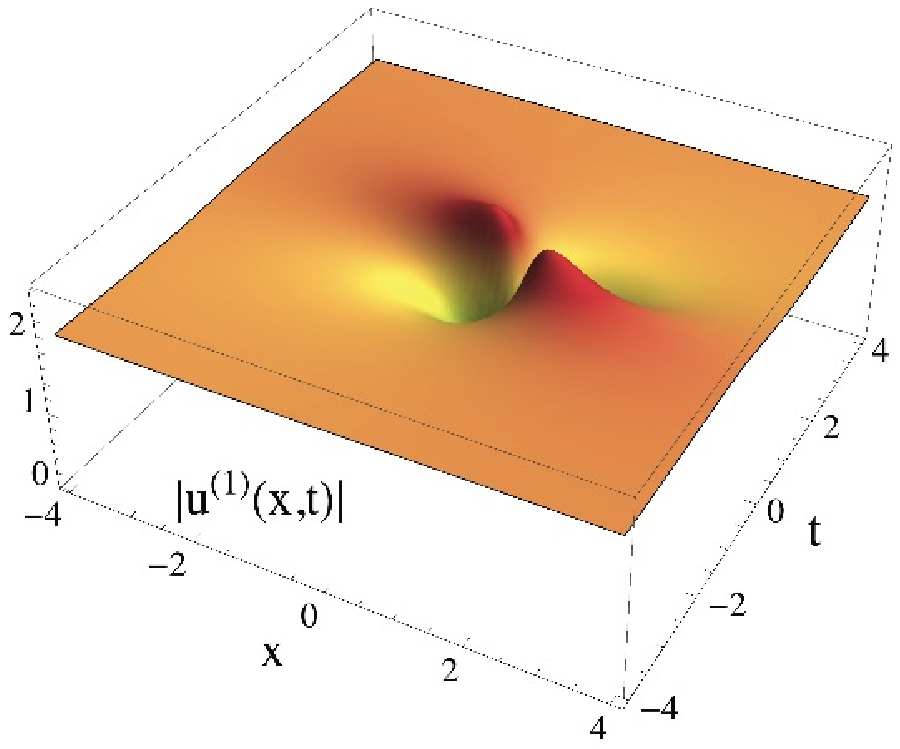}
 %{422a_VNLS_g2_u1}
%\hspace{0.09cm}
 \includegraphics[scale=0.8]{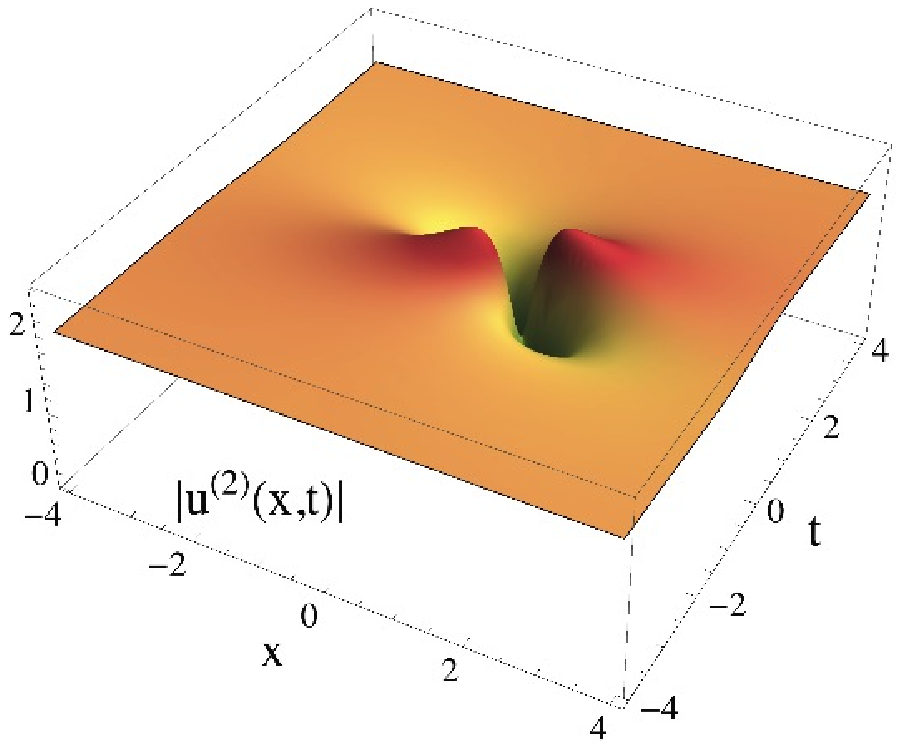}
 %{422a_VNLS_g2_u2}
 %\caption{VNLS:  $k_c=\frac{i}{2}\sqrt{-13+16\sqrt{2}}=i 1.551\,,\,\lambda_1=\,\lambda_2\neq\lambda_3\,,s_1= s_2=1$, $q=1$, $a_1=a_2=2$,  $\epsilon=1$;  $\gamma_2 =1$, $\gamma_1 =\gamma_3 =0$.}
 \caption{VNLS:  $k_c=\frac{i}{2}\sqrt{-13+16\sqrt{2}}$, $\lambda_1=\lambda_2\neq\lambda_3$, $s_1= s_2=1$, $q=1$, $a_1=a_2=2$;  $\gamma_2 =1$, $\gamma_1 =\gamma_3 =0$.}
 \label{figure6}
 \end{centering} 
 \end{figure}

\begin{figure}[H]
 \begin{centering} 
 \includegraphics[scale=0.8]{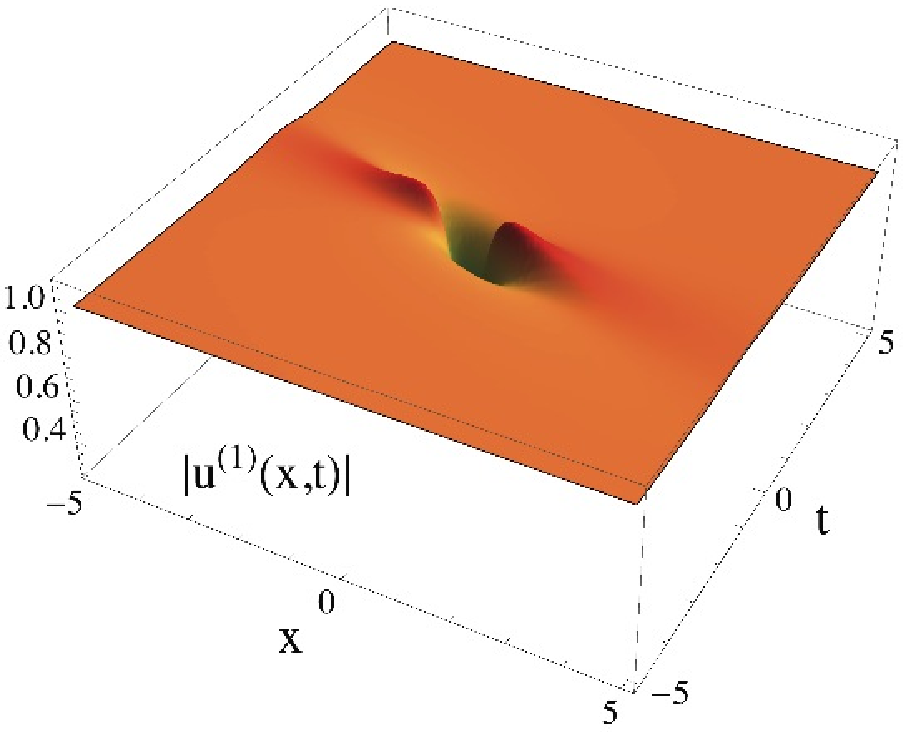}
 %{422b_VNLS_g2_u1}
%\hspace{0.09cm}
 \includegraphics[scale=0.8]{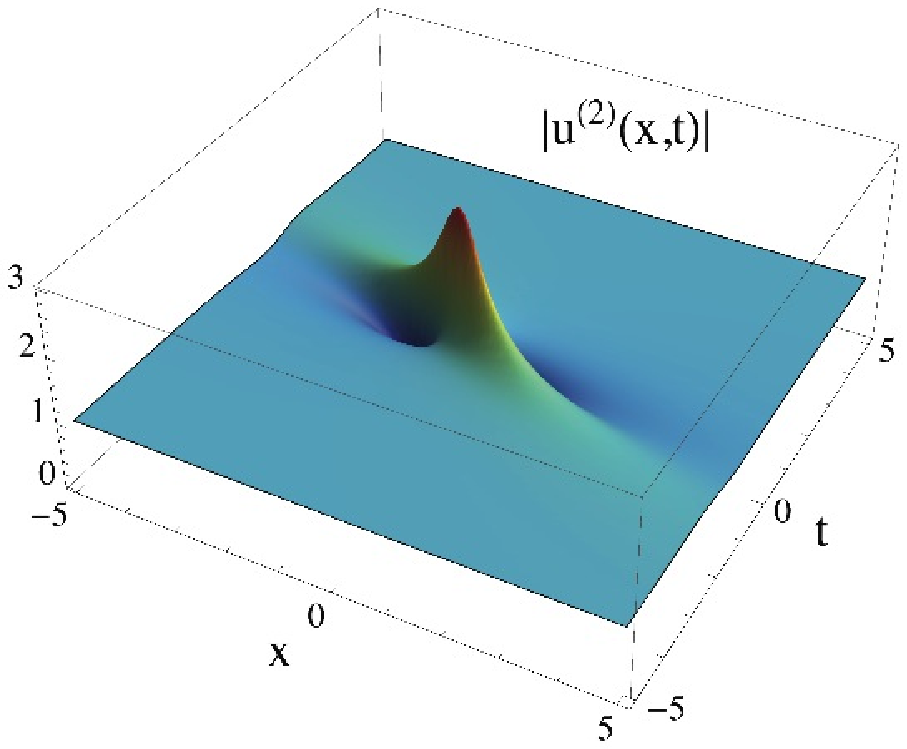}
 %{422b_VNLS_g2_u2}
 \caption{VNLS:  $k_c=\sqrt{\frac38} \sqrt{-3+i\sqrt{3}}$, $\lambda_1=\lambda_2\neq\lambda_3$, $s_1= s_2=-1$, $q=1$, $a_1=a_2=1$;  $\gamma_2 =1$, $\gamma_1 =\gamma_3 =0$.}
 \label{figure7}
 \end{centering} 
 \end{figure}

\begin{figure}[H]
 \begin{centering} 
 \includegraphics[scale=0.8]{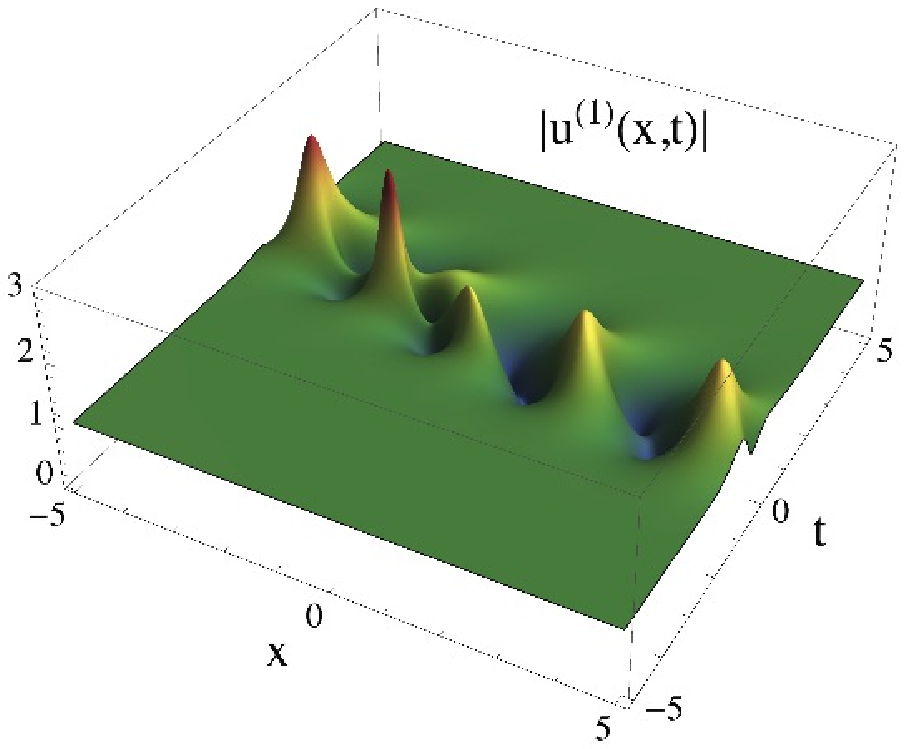}
 %{422b_g3_figure8a} 
%\hspace{0.09cm}
 \includegraphics[scale=0.8]{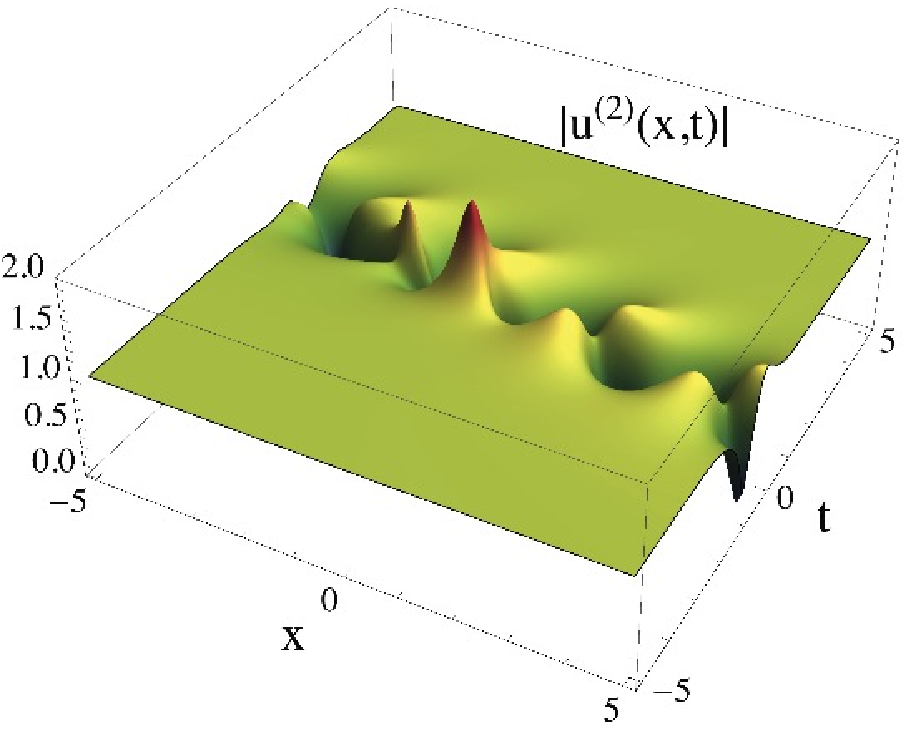}
 %{422b_g3_figure8b}
 \caption{VNLS:  $k_c=\sqrt{\frac38} \sqrt{-3+i\sqrt{3}}$, $\lambda_1=\lambda_2\neq\lambda_3$, $s_1= s_2=-1,q=1$, $a_1= a_2=1$;  $\gamma_1=\gamma_2 =\gamma_3 =1$.}
 \label{figure8}
 \end{centering} 
 \end{figure}

\begin{figure}[H]
 \begin{centering} 
 \includegraphics[scale=0.53]{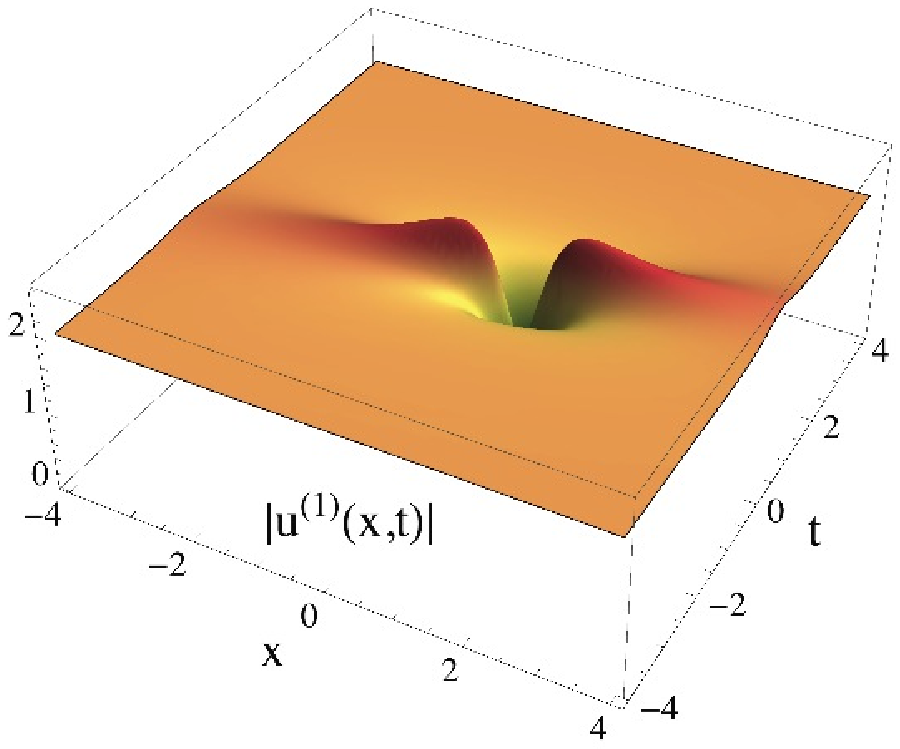} 
%\hspace{0.09cm}
 \includegraphics[scale=0.53]{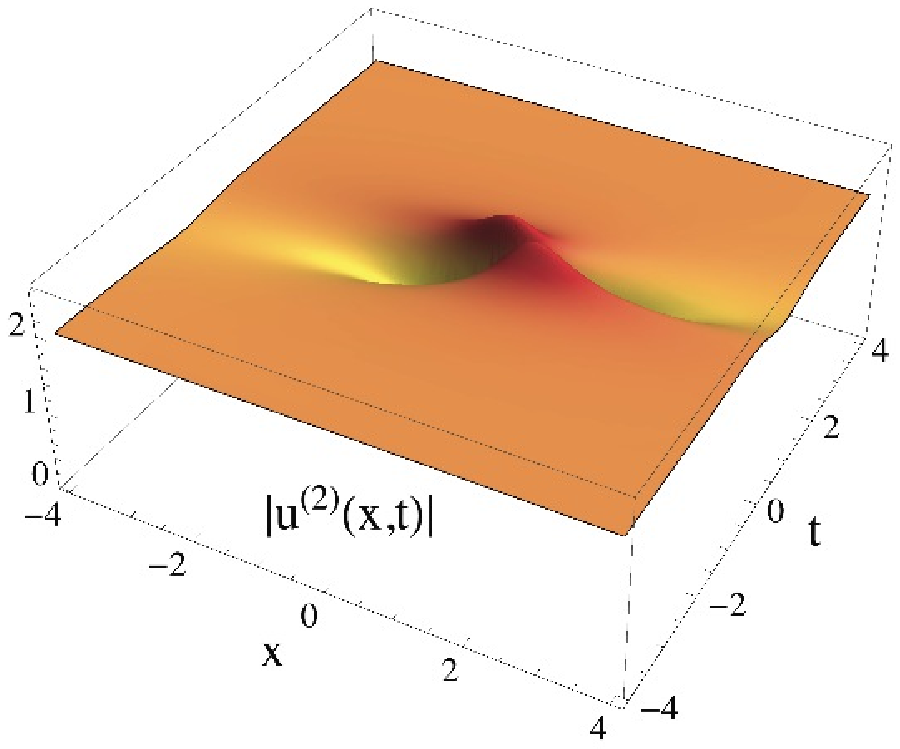}
% \hspace{0.09cm}
 \includegraphics[scale=0.53]{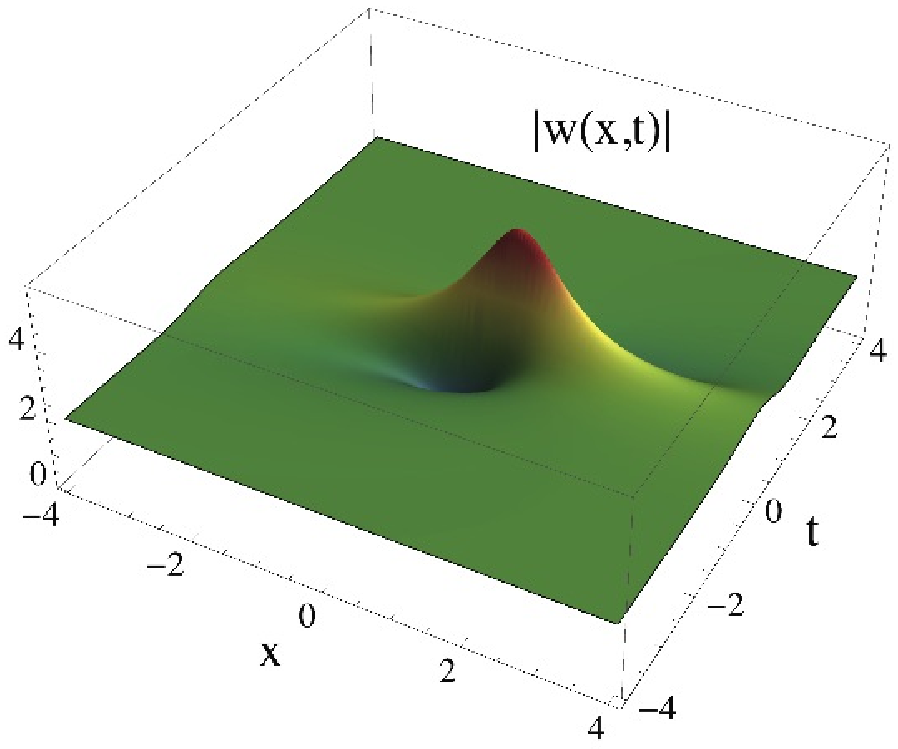}
 \caption{3WRI:  $k_c=\frac{i}{2}\sqrt{-13+16\sqrt{2}}$, $\lambda_1=\lambda_2\neq\lambda_3$, $s_1= s_2=1\,,\,q=1$, $a_1= a_2=2$, $c_1=1$, $c_2=2$;  $\gamma_2=1$, $\gamma_1 =\gamma_3 =0$.}
 %; max value .} 
 \label{figure9}
 \end{centering} 
 \end{figure}

\begin{figure}[H]
 \begin{centering} 
 \includegraphics[scale=0.53]{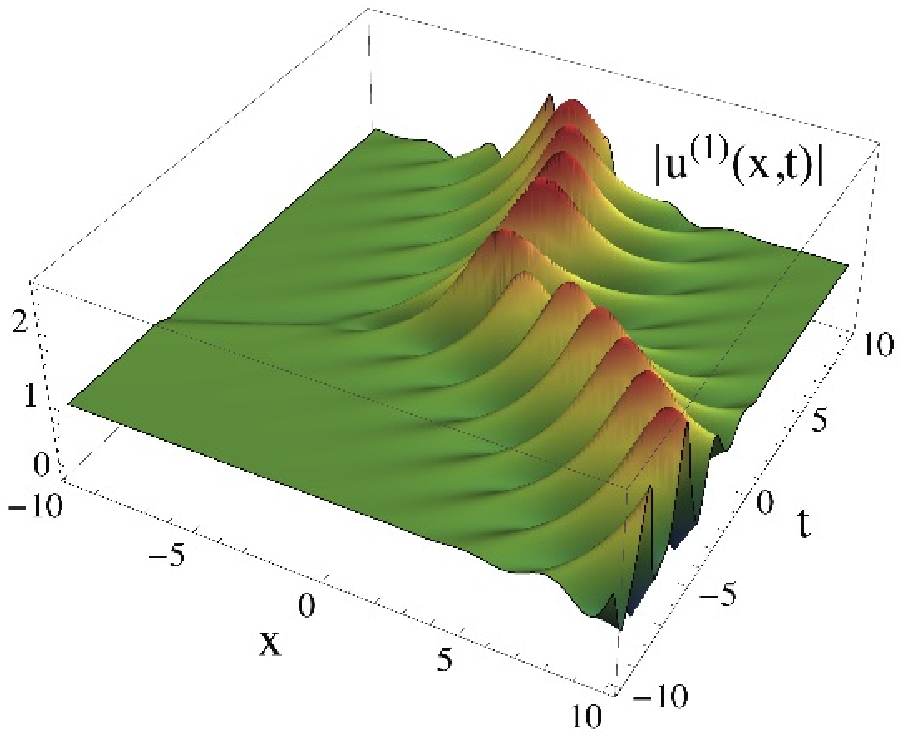} 
%\hspace{0.09cm}
 \includegraphics[scale=0.53]{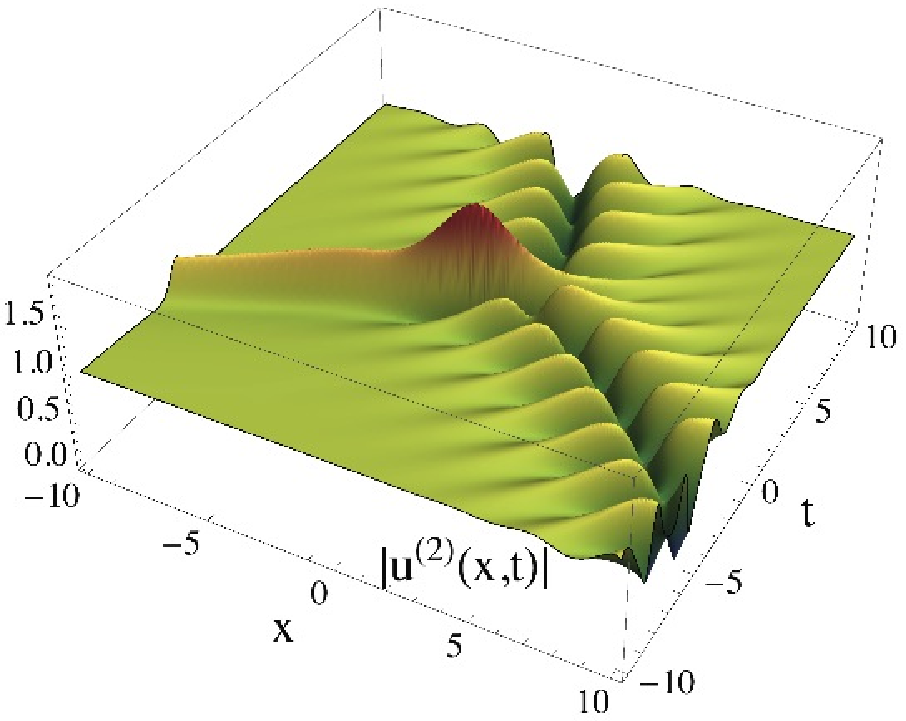}
% \hspace{0.09cm}
 \includegraphics[scale=0.53]{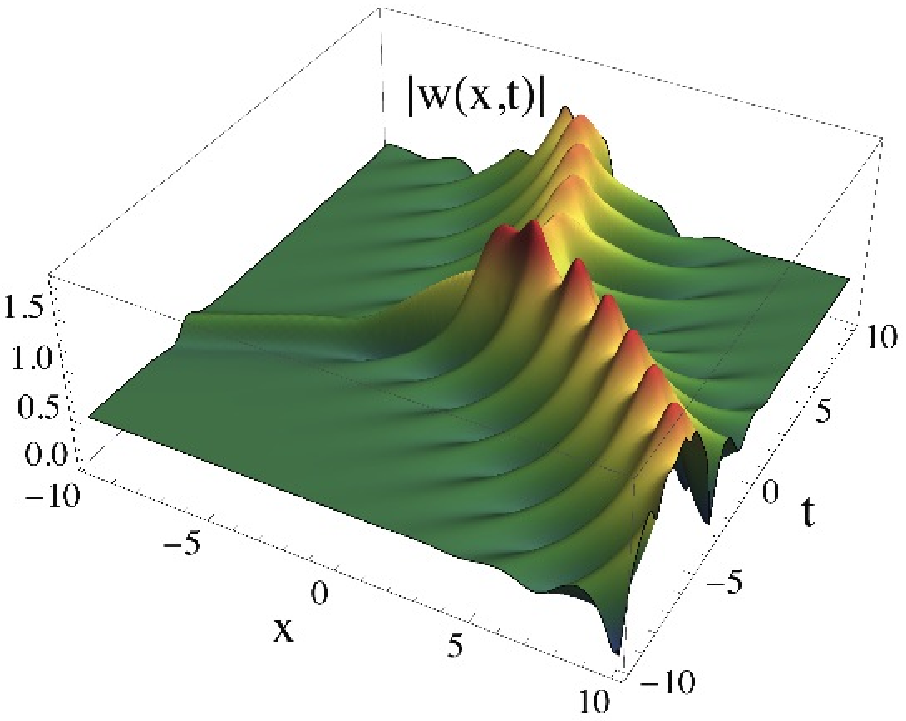}
 \caption{3WRI:  $k_c=\sqrt{\frac38} \sqrt{-3+i\sqrt{3}}$, $\lambda_1=\lambda_2\neq\lambda_3$, $s_1= s_2=-1$, $q=1$, $a_1= a_2=1$, $c_1=1$, $c_2=2$;   $\gamma_1 =\gamma_2=\gamma_3 =1$.}
 %; max value .} 
 \label{figure10}
 \end{centering} 
 \end{figure}

This family of solutions possesses two novel features with respect to those discussed in the previous subsections. First, the choice $s_1=s_2=1$ is compatible with the boundedness of solutions (see below). Second, the  conditions on the parameter set for the existence of a critical value $k_c$ lead to threshold phenomena for the dimensionless positive parameter $m=a_1^2/q^2$. As implied by the explicit expression (\ref{discrzero}) of the zeros of the discriminant (\ref{discri}), alias (\ref{discri2}), we state the following 
 \begin{Proposition}
Assume $s_1=s_2=1$:
\begin{enumerate}
\item if $q^2\,\geq\,2a_1^2$  then the four zeros $k(\eta_1,\eta_2)$, see (\ref{discrzero}),  are real and no (complex) critical value $k_c$ exists.
\item if $q^2\,<\,2a_1^2$ then the two zeros $k(1,\eta_2)$ are real and the other two $k(-1,\eta_2)$ are imaginary. Therefore in this subset of the parameter plane $(a_1\,,\,q)$  there are two critical values with opposite sign, i.e. $k_c=k(-1,\eta_2)$ or, explicitly,
\begin{equation}\label{kcs=1}
k_c=k(-1,\eta_2)= i\eta_2\left( \frac12 D_2+\sqrt{\frac14 D_2^2-D_0}\right)^{1/2},\quad \eta_2^2=1\,,
\end{equation}
where $D_0$ and $D_2$ are given by (\ref{solvcoeff}) with $s=1$.
\end{enumerate}
\end{Proposition}
\begin{Proposition}
 Assume $s_1=s_2=-1$:
\begin{enumerate} 
\item if $q^2\,>\,\frac{1}{4}a_1^2$ then the four zeros $k(\eta_1,\eta_2)$, see (\ref{discrzero}), are strictly complex (namely Im$[k]\neq 0$) and therefore there are four critical values $k_c=k(\eta_1,\eta_2)$.
\item if $q^2\,\leq\,\frac{1}{4}a_1^2$ then the four zeros are imaginary and the critical values are again $k_c=k(\eta_1,\eta_2)$.
\end{enumerate}
\end{Proposition}
Once $k_c$ is computed, 
%the construction of the solution $u^{(1)}\;,\;u^{(2)}\;,\;w$ goes via the following chain of steps:
its corresponding solution of the equations (\ref{3W-NLS}) is obtained through the following chain of steps:
i) use Proposition 2 to compute the eigenvalues $\lambda_1$ and $\lambda_3$, ii) compute $\omega_1$, $\omega_3$, $\rho$ according to (\ref{omegaexpr}), iii) insert the expression (\ref{Tfine}) of the similarity matrix $T$ in (\ref{zcritrat2}) to compute the vector $V$, iv) finally apply the Darboux-Dressing formula (\ref{NEW}).
 Instances of solutions of the VNLS equation are shown in Figure \ref{figure6} (rational, defocusing), Figure \ref{figure7} (rational, focusing), Figure \ref{figure8} (rational-exponential, focusing). Instances of solutions of the 3WRI equation are shown in Figure \ref{figure9} (rational, $s_1=s_2=1$), Figure \ref{figure10} (rational-exponential, $s_1=s_2=-1$).

 %%%%%%%%%%%%%%%% figure 6 %%%%%%%%%%%%%%%%%%%
 
 %%%%%%%%%%%%%%%% figure 7 %%%%%%%%%%%%%%%%%%%
 
  %%%%%%%%%%%%%%%% figure 8 %%%%%%%%%%%%%%%%%%%
 
%%%%%%%%%%%%%%%%%% 3WRI  si=1 %%%%%%%%%%%%%%%%%%%  
 
%%%%%%%%%%%%%%%%%% si=-1  - quale?  %%%%%%%%%%%%%%%%%%%

%\begin{figure}[H]
% \begin{centering} 
% \includegraphics[scale=0.53]{422b_3WRI_g3_u1} 
%%\hspace{0.09cm}
% \includegraphics[scale=0.53]{422b_3WRI_g3_u2}
%% \hspace{0.09cm}
% \includegraphics[scale=0.53]{422b_3WRI_g3_w}
% \caption{3WRI:  $ k_c=\sqrt{\frac38} \sqrt{-3+i\sqrt{3}}$, $\lambda_1=\,\lambda_2\neq\lambda_3\,,\,s_1= s_2=-1\,,\,q=1$, $a_1= a_2=1$, $c_1=1$, $c_2=2$;  $\gamma_2=0$, $\gamma_1 =\gamma_3 =1$.}
% %; max value .} 
% \label{figure9}
% \end{centering} 
% \end{figure}

 %%%%%%%%%%%%%%SUBSECTION 4.2.3%%%%%%%%%%%

\subsubsection{$q\neq 0$ and $a_1\neq a_2$}

\begin{figure}[H]
 \begin{centering} 
 \includegraphics[scale=0.8]{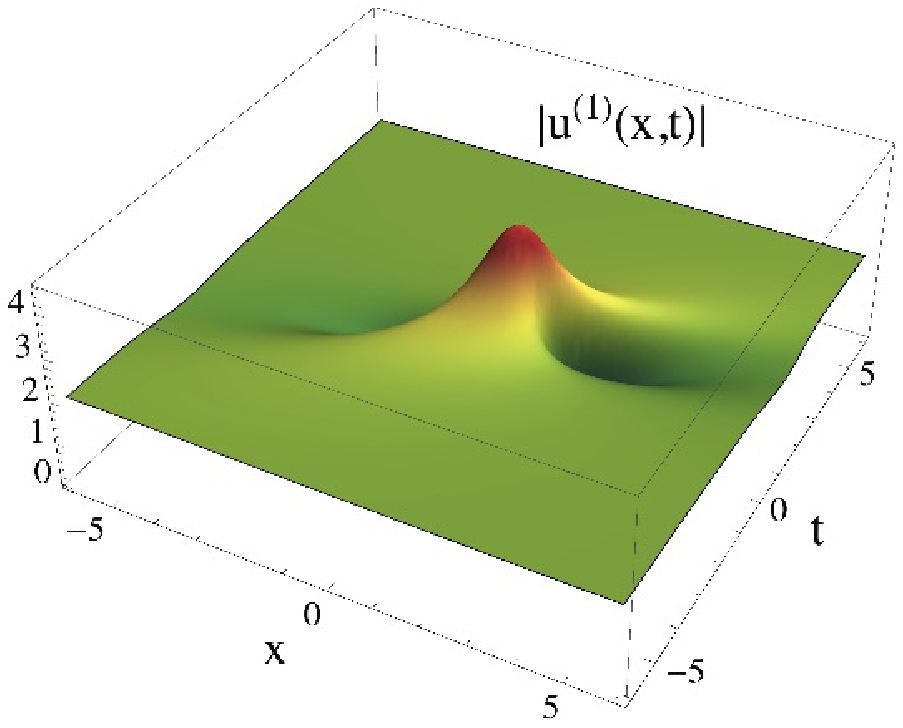} 
%\hspace{0.09cm}
 \includegraphics[scale=0.8]{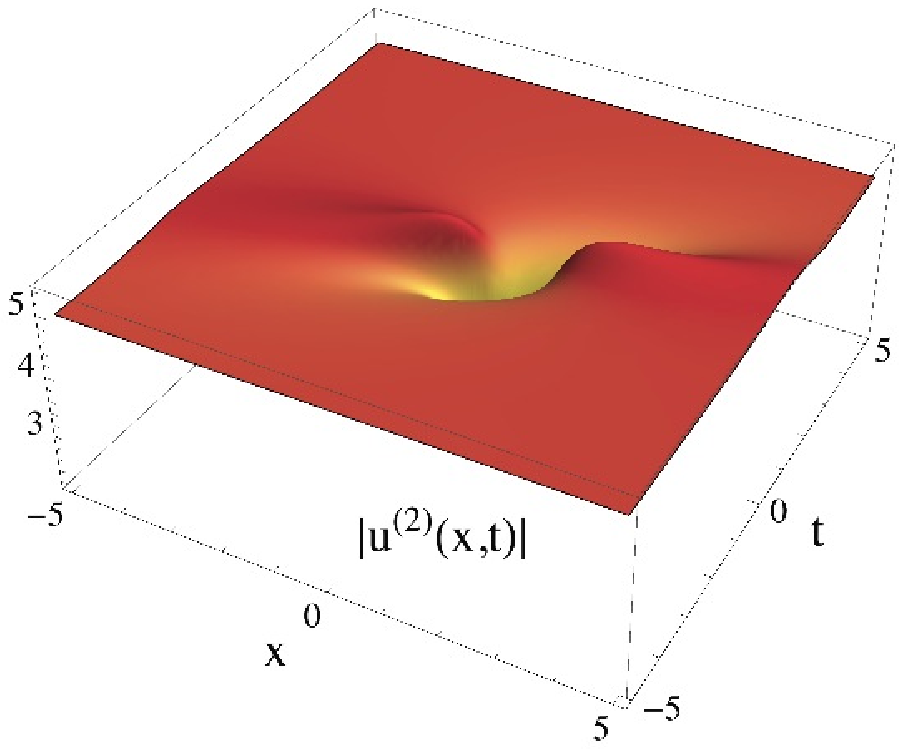}
 \caption{VNLS:  $k_c=-5.600+4.655 i$, $\lambda_1=\lambda_2\neq\lambda_3$, $s_1= s_2=1$, $q=1$, $a_1=2$, $a_2=5$;  $\gamma_2 =1$, $\gamma_1 =\gamma_3 =0$.}
 %; max value .} 
 \label{figure11}
 \end{centering} 
 \end{figure}

\begin{figure}[H]
 \begin{centering} 
 \includegraphics[scale=0.8]{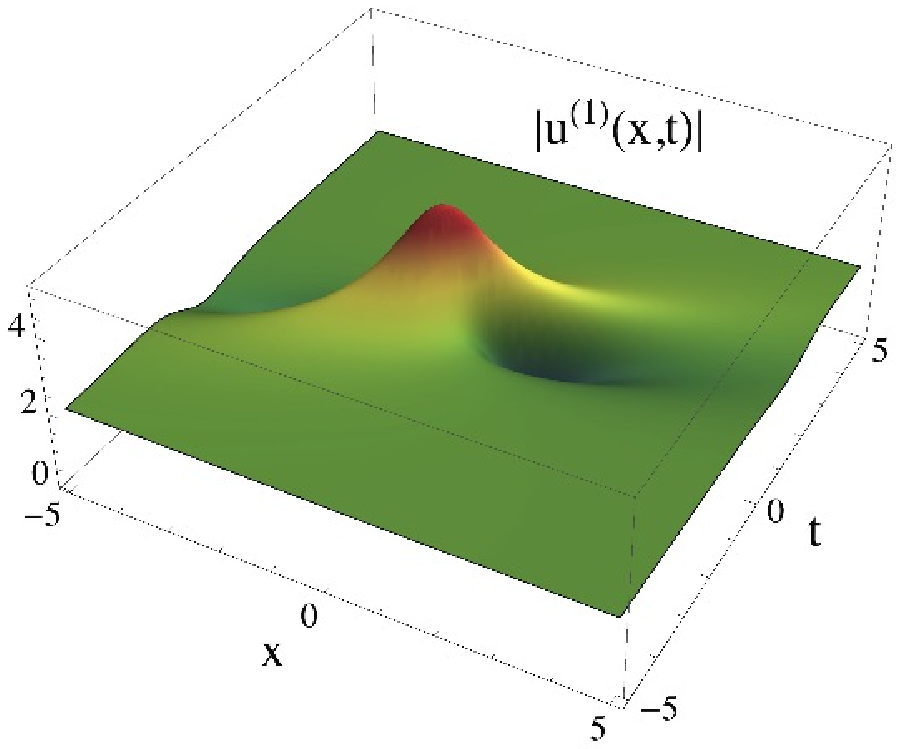} 
%\hspace{0.09cm}
 \includegraphics[scale=0.8]{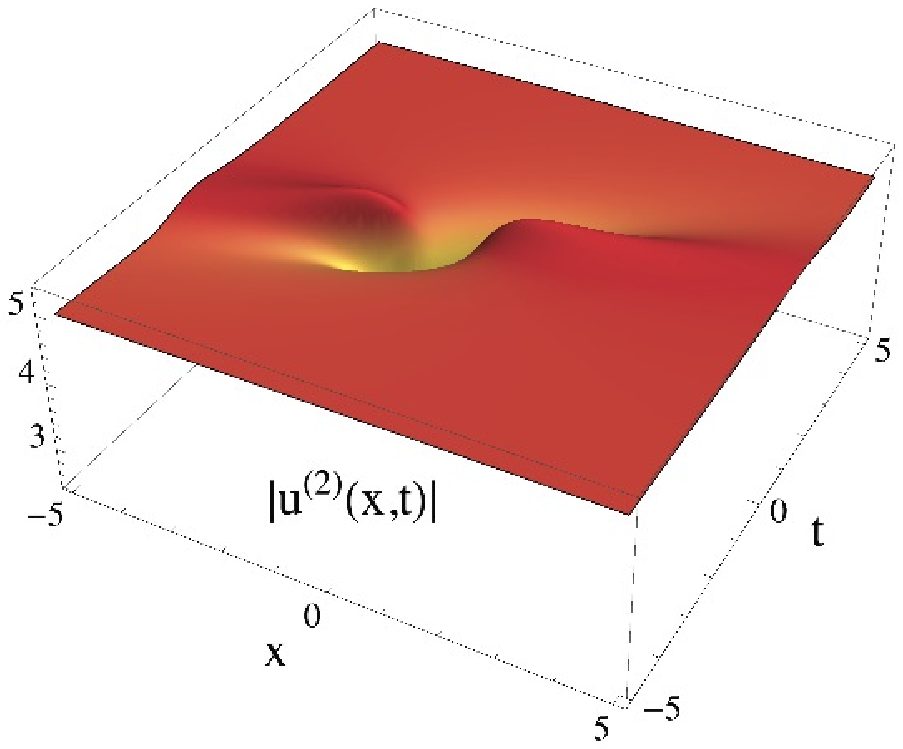}
 \caption{VNLS: $k_c=4.876+5.343 i$, $\lambda_1=\lambda_2\neq\lambda_3$, $s_1= s_2=-1$, $q=1$, $a_1=2$, $a_2=5$;  $\gamma_2 =1$, $\gamma_1 =\gamma_3 =0$.}
 %; max value .} 
 \label{figure12}
 \end{centering} 
 \end{figure}

\begin{figure}[H]
 \begin{centering} 
\includegraphics[scale=0.8]{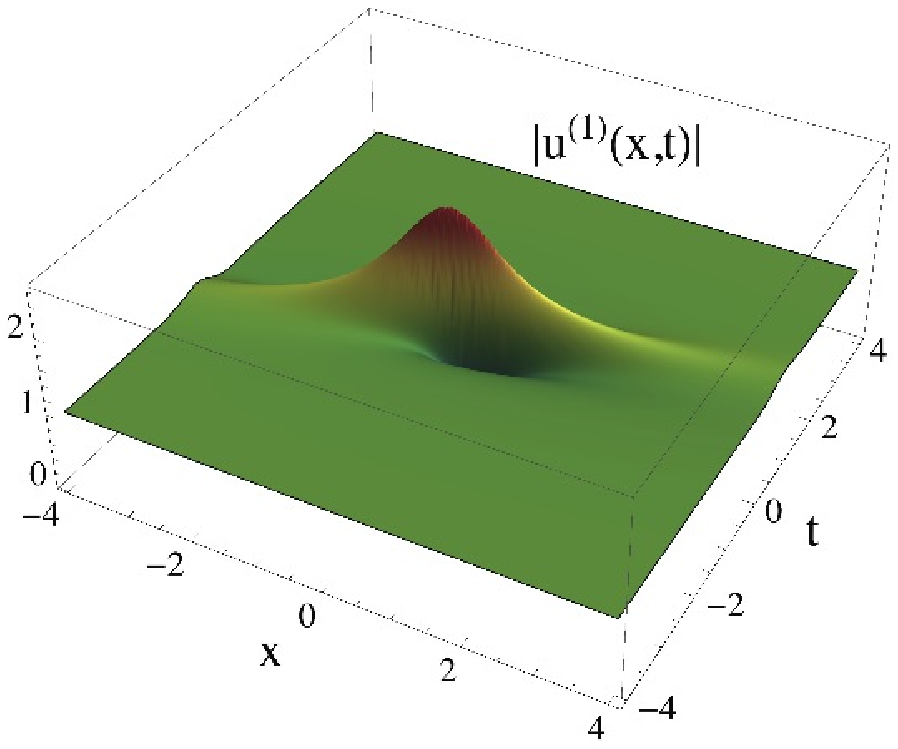} 
%\hspace{0.09cm}
\includegraphics[scale=0.8]{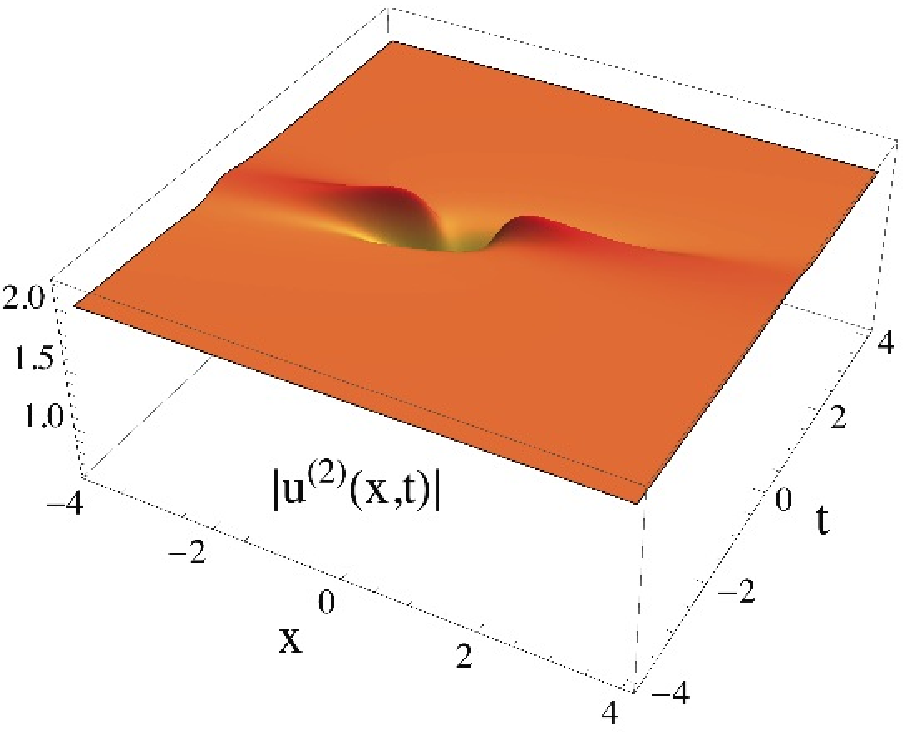}
\caption{VNLS: $k_c=-1.242+0.636 i$, $\lambda_1=\lambda_2\neq\lambda_3$, $s_1=-1$, $ s_2=1$, $q=1$, $a_1=1$, $a_2=2$;  $\gamma_2 =1$, $\gamma_1 =\gamma_3 =0$.}
 %; max value .} 
 \label{figure13}
 \end{centering} 
 \end{figure}

\begin{figure}[H]
 \begin{centering} 
 \includegraphics[scale=0.8]{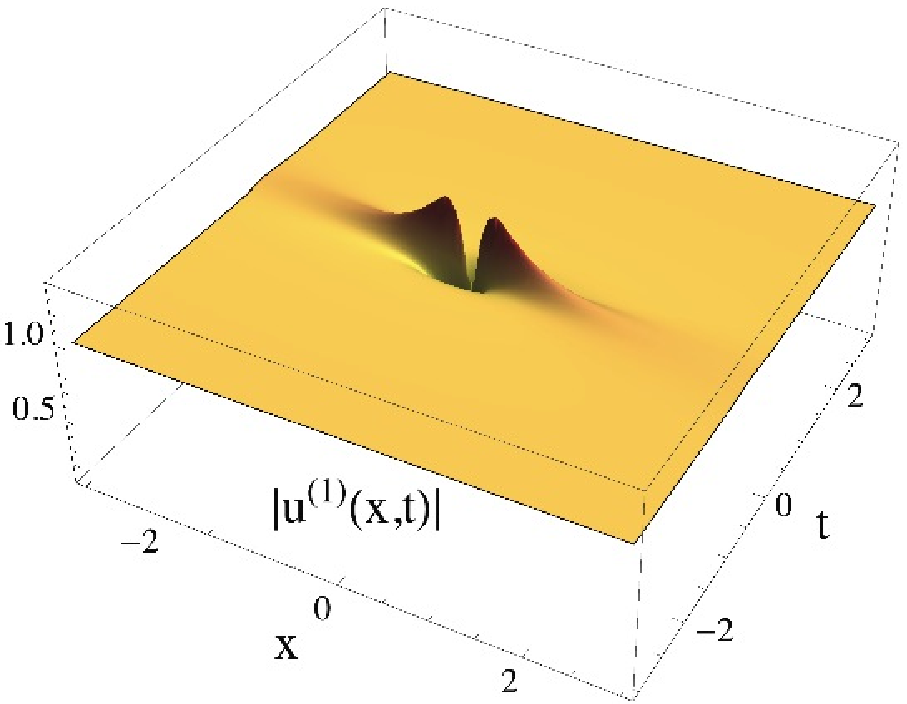} 
%\hspace{0.09cm}
 \includegraphics[scale=0.8]{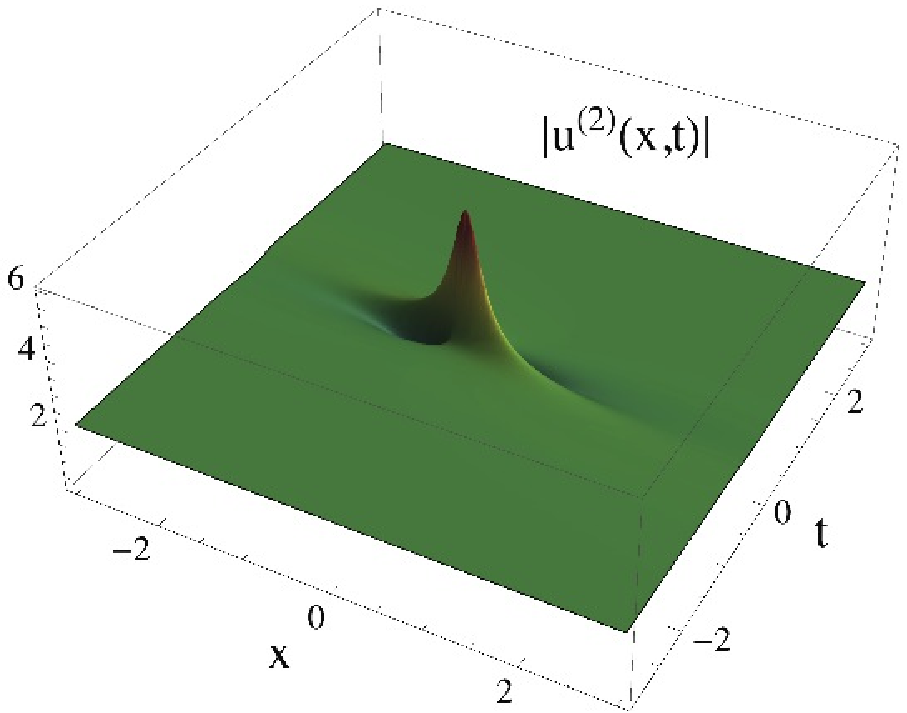}
 \caption{VNLS: $k_c=0.625+1.879 i$, $\lambda_1=\lambda_2\neq\lambda_3$, $s_1=1$, $s_2=-1$, $q=1$, $a_1=1$, $a_2=2$;  $\gamma_2 =1$, $\gamma_1 =\gamma_3 =0$.}
 %; max value .} 
 \label{figure14}
 \end{centering} 
 \end{figure}

\begin{figure}[H]
 \begin{centering} 
 \includegraphics[scale=0.53]{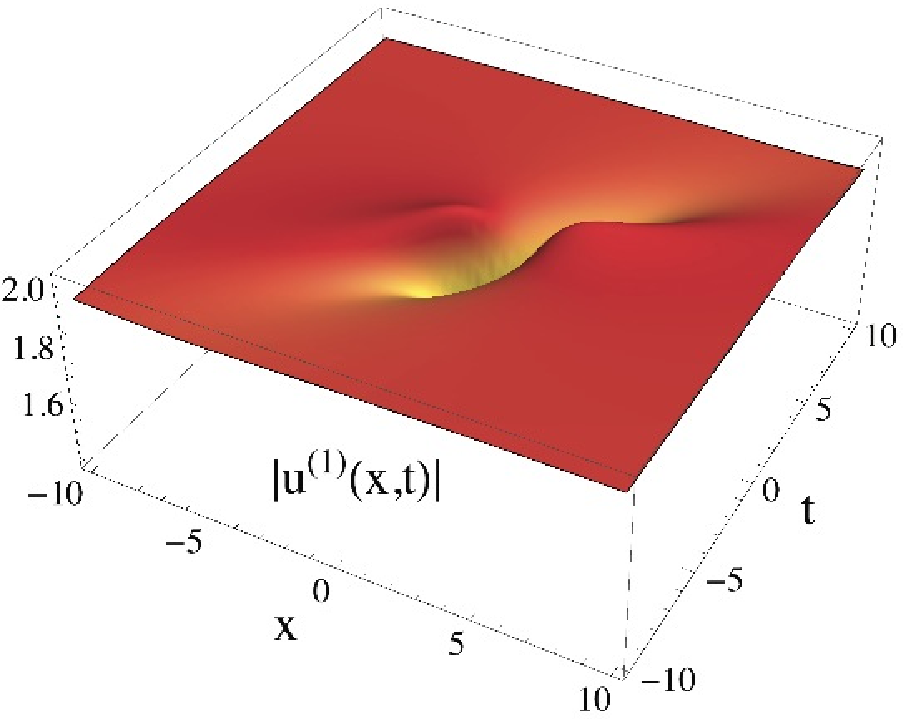} 
%\hspace{0.09cm}
 \includegraphics[scale=0.53]{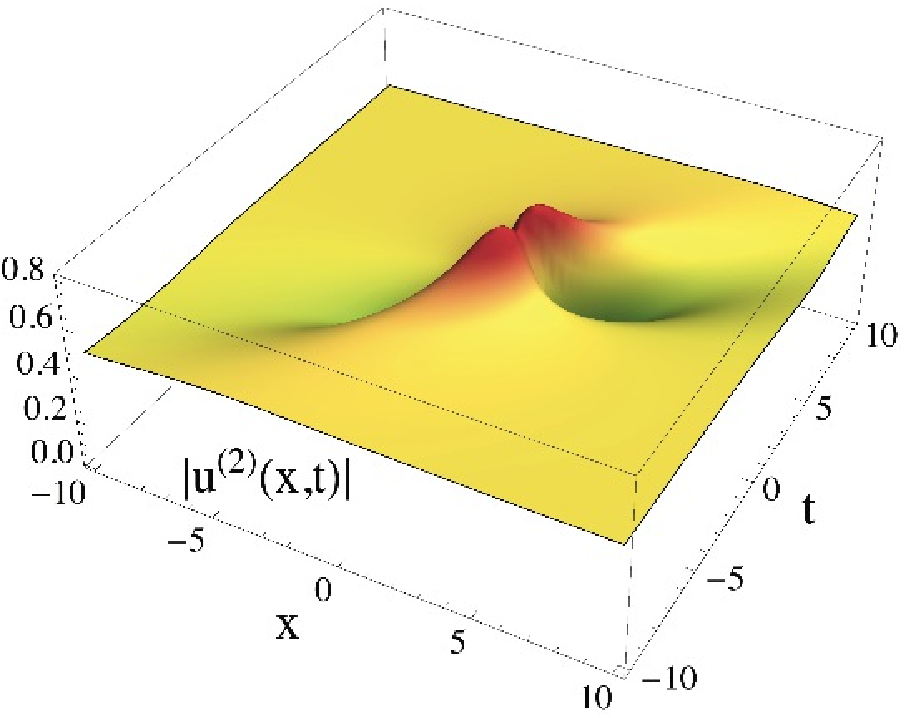}
% \hspace{0.09cm}
 \includegraphics[scale=0.53]{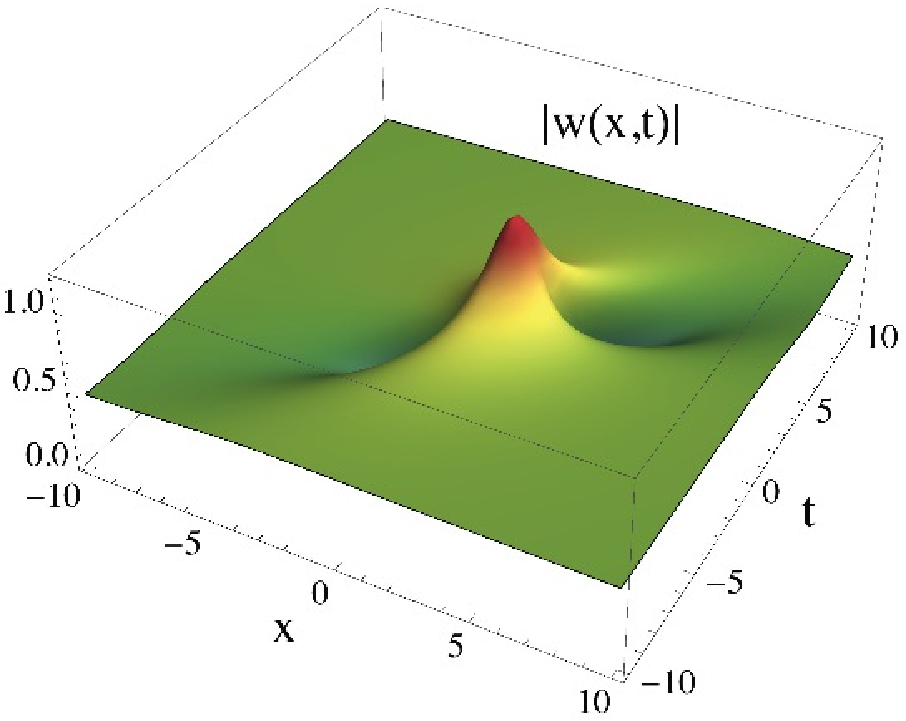}
 \caption{3WRI: $k_c=1.319+0.256 i$,  $\lambda_1=\lambda_2\neq\lambda_3$, $s_1= s_2=1$, $q=1$, $a_1=2$, $a_2=0.5$,  $c_1=1$, $c_2=2$; $\gamma_2=1$, $\gamma_1 =\gamma_3 =0$.}
 %; max value .} 
 \label{figure15}
 \end{centering} 
 \end{figure}

We explore this case by first computing $k_c$ numerically. Then the step-by-step method of construction of the solution, as indicated in the previous subsection, produces the plots of solutions of the VNLS as displayed in Figure \ref{figure11} (rational, defocusing), Figure \ref{figure12} (rational, focusing), Figure \ref{figure13} (rational, $s_1=-1, s_2=1$), Figure \ref{figure14} (rational, $s_1=1, s_2=-1$). An instance of solution of the 3WRI equation is shown in Figure \ref{figure15} (rational, $s_1=s_2=1$).

%%%%%%%%%%%%%%%%%% VNLS - si=1 %%%%%%%%%%%%

 %%%%%%%%%%%%%%%%%% VNLS - si=-1 %%%%%%%%%%%%
 
  %%%%%%%%%%%%%%%%%%%%%%%%% SEGNI MISTI %%%%%%%%%%%%%%%%% 
 
 %%%%%%%%%%%%%%%% subsection 4.3 %%%%%%%%%%%%%%%%%%%%%%%%
 
 \subsection{Conclusions}
 \label{conclusions}
 In this article we have devised a method of construction of solutions of two integrable systems of partial differential equations of interest in a variety of applications. These systems,  the VNLS equations and the 3WRI equations, model the coupling of two waves and, respectively, of three waves. Our construction is specially tailored to yield solutions which feature a rational, or mixed rational-exponential, dependence on the independent variables. 
While rational solutions of integrable partial differential equations attracted mathematical interest since the 70's and consequently  this type of solutions were derived for a number of integrable wave equations, it was only recently that further investigations of rational solutions were extended to integrable systems of two or three coupled differential equations. The main motivation of such a renewed interest goes back to the observation by Peregrine that the simplest rational solution of the focusing NLS equation may well model an ocean rogue wave. In a variety of physical contexts it was however soon recognized that, several waves, rather than a single one, should be considered in order to account for important resonant interaction processes. 
%This type of solutions have been derived since decades for a number of other integrable wave equations. 
For integrable partial differential equations, according to personal taste, various, yet equivalent, approaches have been adopted: spectral transform and dressing techniques, Wronskian and Hirota methods, and Darboux transformations as considered here. 
These solutions are all soliton solutions since their corresponding spectral data on the continuos spectrum vanish. Moreover the strategy of computation may depend on whether the soliton is superimposed to the vacuum (i.e. the vanishing solution) or to a plane wave background. Here we deal with this second type of solitons. In most of the constructions discussed in the literature, the way to obtain polynomials out of (a linear combination of) exponentials goes through an appropriate limit process by making a number of eigenvalues of the Lax equations coalesce to get all the same value. Our approach is instead based on the exponentiation of non diagonalizable matrices. This construction naturally leads to consider those critical values $k_c$ of the spectral variable $k$ such that the matrices which appear as exponent are similar to a Jordan form. There is therefore no need to take the limit in which different eigenvalues coalesce. We believe that our investigation is able to capture all possible solutions in this class. We are confident that the broad family of solutions presented here add a contribution to the understanding of rogue wave phenomena in novel physical situations where wave resonant interactions are relevant.

%\newpage
\appendix
\section{Polynomials}
\label{ap:A}
%%%%%%%%% VNLS Poly %%%%%%%%%%%%%%%%%%%%%%%%%%%%%%%%%%

\begin{subequations}\label{gamma3VNLSpolyn}
{\small \begin{align}
\label{P14}
&P^{(1)}_4=\nonumber\\
&12 X^4+1728 T^4+288 X^2 T^2+ 4\left(3 i+2 \sqrt{3} \epsilon \right)X^3-864 i T^3-72 i X^2 T +48\,X T^2\left(3 i-2 \sqrt{3}\epsilon\right) +\nonumber\\
& + 3\,X^2\left(4 \text{Re}[\gamma_1] +3 i \sqrt{3}\epsilon -1\right)-12\,T^2\left(12\text{Re}[\gamma_1]+5 i \sqrt{3} \epsilon +9\right) +\nonumber\\
&+  12\,T X \left(4\sqrt{3} \epsilon \text{Im}[\gamma_1]  - i \sqrt{3} \epsilon +9\right)
+6\,T\left(2 i \sqrt{3} \epsilon \text{Im}[\gamma_1]  +6 i \text{Re}[\gamma_1]- \sqrt{3}\epsilon +3 i\right)
+\nonumber\\
& +2\, X\left(-3 i \sqrt{3} \epsilon \text{Im}[\gamma_1] +2 \sqrt{3}\epsilon \text{Re}[\gamma_1]+3 i \text{Re}[\gamma_1]-2 \sqrt{3} \epsilon -3 i\right)  +3 |\gamma_1|^2+ \nonumber\\
& +\frac{1}{2}\left(1+5 i \sqrt{3}\epsilon\right)\text{Re}[\gamma_1]+\frac{9}{2}\left(\sqrt{3}  \epsilon - i\right) \text{Im}[\gamma_1]+\frac{5}{2}\left(1-i \sqrt{3} \epsilon\right)
\end{align} 
\begin{align}
\label{P24}
&P^{(2)}_4=\nonumber\\
&12 X^4+ 1728 T^4+288 X^2 T^2+4\left(-3i+2 \sqrt{3} \epsilon\right)X^3 -864 i T^3-72 i X^2 T - 48 \,XT^2\left(3i+2 \sqrt{3}\epsilon \right) +\nonumber\\
&+3\, X^2\left(4 \text{Re}[\gamma_1] -3 i \sqrt{3} \epsilon -1\right) -12\, T^2\left(12
\text{Re}[\gamma_1] -5 i \sqrt{3} \epsilon +9\right) +\nonumber\\
&+12\, T X \left(4\sqrt{3}\epsilon \text{Im}[\gamma_1]  - i \sqrt{3} \epsilon +3\right)+ 6\, T\left(-2 i \sqrt{3}\epsilon \text{Im}[\gamma_1]   +6 i \text{Re}[\gamma_1] + \sqrt{3}\epsilon -3 i\right)
+\nonumber\\
& +2\,X\left(-3 i \sqrt{3} \epsilon \text{Im}[\gamma_1]  +2 \sqrt{3} \epsilon \text{Re}[\gamma_1]  -3 i \text{Re}[\gamma_1] -2 \sqrt{3} \epsilon -6 i\right) +3 |\gamma_1|^2+ \nonumber\\
&+ \frac{1}{2}\left( 1-5 i \sqrt{3} \epsilon \right)\text{Re}[\gamma_1] +\frac{3}{2}\left( \sqrt{3} \epsilon-3i\right)\text{Im}[\gamma_1] -2\left(1+ i \sqrt{3} \epsilon\right) 
\end{align} }
\begin{align}
\label{P4}
& P_4 =12 X^4+1728 T^4+288 X^2 T^2 +8\sqrt{3} \epsilon X^3-96 \sqrt{3}\epsilon X T^2 +6X^2(1+2\text{Re}[\gamma_1] ) +\nonumber\\
&+72 T^2(1-2 \text{Re}[\gamma_1] )+12 X T( 6+4 \sqrt{3}\epsilon \text{Im}[\gamma_1])  +2X \sqrt{3} \epsilon( 1+2 \text{Re}[\gamma_1])  +\nonumber\\
  &+3 |\gamma_1|^2 - \text{Re}[\gamma_1] +3 \sqrt{3}\epsilon \text{Im}[\gamma_1]  +4
\end{align} 
\end{subequations}

%%%%%%% 3WRI Poly %%%%%%%%%%%%%%%%%%%%%%%%%%%%%%%%%%%

\begin{subequations}\label{gamma23Wpolyn}
\begin{align}\label{Q12}
Q^{(1)}_2&=12 X^2+12 T^2 \left(c_1^2-c_1 c_2+c_2^2\right) -12 X T  (c_1+c_2) + 2 X\left(2 \sqrt{3} \epsilon +3 i\right)+\nonumber\\
&-2  \,T\left[c_1 \left(\sqrt{3} \epsilon -3 i\right)+c_2 \left(\sqrt{3} \epsilon +6
   i\right)\right] +i \sqrt{3} \epsilon -1
\end{align}
\begin{align}\label{Q22}
Q_2&=12 X^2+12 \,T^2 \left(c_1^2-c_1 c_2+c_2^2\right) -12\,X T  (c_1+c_2) +4\,X \left(\sqrt{3} \epsilon -3 i\right)+\nonumber\\
&-2 \,T(c_1+c_2)\left(\sqrt{3} \epsilon -3 i\right)  -2 i \sqrt{3} \epsilon -1
\end{align}
\begin{equation}\label{M2}
M_2=12 X^2+12 T^2 \left(c_1^2-c_1 c_2+c_2^2\right)-12X T  (c_1+c_2)+4 \sqrt{3} \epsilon X-2 \sqrt{3} \epsilon  T (c_1+c_2) +2
\end{equation}
\end{subequations}

\begin{subequations}\label{gamma33Wpolyn}
{\small \begin{align}
\label{Q14}
&Q^{(1)}_{4} =
12X^4+12 T^4 \left(c_1^{2}-c_1 c_2+c_2^{2}\right)^2+36 X^2 T^2 \left(c_1^{2}+c_2^{2}\right)-24 X^3 T  (\text{c1}+\text{c2})-24 X T^3  \left(c_1^{3}+c_2^{3}\right)+
  \nonumber\\ 
 &+4 X^3 \left(2 \sqrt{3} \epsilon +3 i\right)+  \nonumber\\ 
 &-4 T^3 \left[c_1^{3} \left(4\sqrt{3} \epsilon -3 i\right)-3 c_1^{2} \text{c2}
   \left(\sqrt{3} \epsilon -3 i\right)-3 c_1c_2^{2} \left(\sqrt{3} \epsilon
   +3 i\right)+ 2c_2^{3} \left(2\sqrt{3} \epsilon +3 i\right)\right]+
    \nonumber\\  
&-12 X^2 T  \left[\sqrt{3} \epsilon c_1+c_2\left(\sqrt{3} \epsilon +3 i\right)\right]   +12 X T^2  \left[3 \sqrt{3} \epsilon c_1^{2}  -4 \sqrt{3} \epsilon c_1c_2
   +3  c_2^{2} \left(\sqrt{3}\epsilon  +i\right)\right]+
    \nonumber\\  
& +3 X^2 \left(4 \text{Re}[\gamma_1] +3 i \sqrt{3} \epsilon -1\right)+
    \nonumber\\  
&-3 T^2 \left[c_1^{2} \left(2\text{Re}[\gamma_1]  -\sqrt{3} \epsilon ( 2\text{Im}[\gamma_1] -i) -11\right)+2 c_1c_2\left(-4 \text{Re}[\gamma_1] +i \sqrt{3} \epsilon+7\right)+  \right. \nonumber\\ 
&+\left. 2c_2^{2} \left(\text{Re}[\gamma_1]+\sqrt{3}\epsilon  (\text{Im}[\gamma_1] -3 i) 
    -1\right)\right]+
    \nonumber\\  
&-6 X T  \left[2c_1\left(\sqrt{3} \epsilon (\text{Im}[\gamma_1] - i)  +
   \text{Re}[\gamma_1] +2\right)+c_2\left(\sqrt{3} \epsilon(-2\text{Im}[\gamma_1] +5i) 
   +2\text{Re}[\gamma_1] -5\right)\right]  +
    \nonumber\\  
&   
   +X \left[\text{Re}[\gamma_1]  \left(4\sqrt{3} \epsilon
   +6 i\right)-2\sqrt{3}\epsilon (2+3 i \text{Im}[\gamma_1] )  -6 i\right]+
     \nonumber\\  
&     
  -2 T \left[c_1\left(\text{Re}[\gamma_1]  \left(\sqrt{3} \epsilon +6 i \right)+9 \text{Im}[\gamma_1] +5
   \sqrt{3} \epsilon -6 i\right)+ \right. \nonumber\\ 
 &+\left.  c_2\left(\text{Re}[\gamma_1]  \left(\sqrt{3} \epsilon -3 i \right)-3\text{Im}[\gamma_1]  \left(3+ i \sqrt{3}
   \epsilon \right)-7 \sqrt{3} \epsilon +3
   i\right)\right]+
     \nonumber\\  
&   
   +3|\gamma_1|^2 +\frac{9}{2} \sqrt{3} \epsilon \text{Im}[\gamma_1]  -\frac{9}{2} i \text{Im}[\gamma_1]+\frac{5}{2} i \sqrt{3} \epsilon\text{Re}[\gamma_1]  +\frac{1}{2}\text{Re}[\gamma_1] -\frac{5}{2}i \sqrt{3} \epsilon +\frac{5}{2}
    \end{align}}
{\small \begin{align}
\label{Q24}
&Q^{(2)}_{4} =12X^4+12 T^4 \left(c_1^{2}-c_1 c_2+c_2^{2}\right)^2+36 T^2 X^2 \left(c_1^{2}+c_2^{2}\right)-24 T  X^3 \left(c_1+c_2\right)-24  T^3  X\left(c_1^{3}+c_2^{3}\right)+
  \nonumber\\ 
 &+4 X^3 \left(2 \sqrt{3} \epsilon -3 i\right)+  \nonumber\\ 
 &-4 T^3 \left[2c_1^{3} \left(2 \sqrt{3} \epsilon -3 i\right)-3 c_1^{2} \text{c2}
   \left(\sqrt{3} \epsilon -3 i\right)-3 c_1c_2^{2} \left(\sqrt{3} \epsilon
   +3 i\right)+ c_2^{3} \left(4 \sqrt{3} \epsilon +3 i\right)\right]+
    \nonumber\\  
&-12 X^2 T  \left[c_1\left(\sqrt{3} \epsilon -3 i\right)+\sqrt{3}\epsilon  c_2\right]   +12 X T^2  \left[3 c_1^{2} \left(\sqrt{3} \epsilon -i\right) -4 \sqrt{3} \epsilon c_1c_2+3 \sqrt{3} \epsilon c_2^{2} \right]+
    \nonumber\\  
& +3 X^2 \left(4 \text{Re}[\gamma_1] -3 i \sqrt{3} \epsilon -1\right)+
    \nonumber\\  
&-3 T^2 \left[2c_1^{2} \left(\text{Re}[\gamma_1]-\sqrt{3} \epsilon  ( \text{Im}[\gamma_1] -3i)  -4\right)-2 c_1c_2\left(4 \text{Re}[\gamma_1] +i \sqrt{3} \epsilon-7\right)+   \right. \nonumber\\ 
&+\left. c_2^{2} \left(2\text{Re}[\gamma_1] +\sqrt{3} \epsilon (2\text{Im}[\gamma_1] - i) 
   -5\right)\right]+
    \nonumber\\  
&-6 X T  \left[c_1\left(2\text{Re}[\gamma_1] +\sqrt{3} \epsilon (2\text{Im}[\gamma_1] -5i) 
   +1\right)+2c_2\left( \text{Re}[\gamma_1] -\sqrt{3}  \epsilon (\text{Im}[\gamma_1] - i) -1\right)\right]  +
    \nonumber\\  
&   
   +2X \left[\text{Re}[\gamma_1]  \left(2\sqrt{3} \epsilon
   -3 i\right)-\sqrt{3}\epsilon (3 i \text{Im}[\gamma_1] +2)  -6 i\right]+
     \nonumber\\  
&     
   +2 T \left[-c_1\left(\text{Re}[\gamma_1]\left(\sqrt{3}\epsilon +3 i \right)+3\text{Im}[\gamma_1]  \left(3-i \sqrt{3}
   \epsilon \right)+ 2 \sqrt{3} \epsilon -12i\right)+ \right. \nonumber\\ 
&+\left. c_2\left( \text{Re}[\gamma_1] \left(-\sqrt{3} \epsilon +6 i \right)+9 \text{Im}[\gamma_1] +4
   \sqrt{3} \epsilon -6 i\right)\right]+
     \nonumber\\  
&   
   +3|\gamma_1|^2+\frac{1}{2}\text{Re}[\gamma_1] -\frac{5}{2} i \sqrt{3} \epsilon\text{Re}[\gamma_1] +\frac{3}{2} \sqrt{3} \epsilon \text{Im}[\gamma_1]   -\frac{9  }{2} i \text{Im}[\gamma_1] -2i \sqrt{3} \epsilon -2
    \end{align}}
{\small \begin{align}
\label{Q4}
&Q_{4} =12X^4+12 T^4 \left(c_1^{2}-c_1 c_2+c_2^{2}\right)^2+36 X^2 T^2 \left(c_1^{2}+c_2^{2}\right)-24 X^3 T  (\text{c1}+\text{c2})-24 X T^3  \left(c_1^{3}+c_2^{3}\right)+
 \nonumber\\  
 &+8 X^3 \left( \sqrt{3} \epsilon -3 i\right)-4(c_1+c_2) T^3 \left[\left( c_1^{2} +c_2^{2} \right)\left(4\sqrt{3} \epsilon -3 i\right)+ c_1c_2\left(-7\sqrt{3} \epsilon
   +3 i\right)\right]+
    \nonumber\\  
&-12(c_1+c_2)  X^2 T  \left(\sqrt{3} \epsilon -3i\right) +12X T^2  \left[3\left( c_1^{2} +c_2^{2} \right)\left(\sqrt{3} \epsilon -i\right) -4\sqrt{3}\epsilon c_1 c_2\right]+ \nonumber\\  
&+6 X^2 \left(2 \text{Re}[\gamma_1] -3 i \sqrt{3} \epsilon -2\right)+
    \nonumber\\  
&-3 T^2 \left[c_1^{2} \left(2\text{Re}[\gamma_1] -\sqrt{3} \epsilon ( 2\text{Im}[\gamma_1] -5i) 
   -11\right)-4 c_1c_2\left(2\text{Re}[\gamma_1] +i \sqrt{3} \epsilon
   -5\right)+ \right. \nonumber\\ 
&+\left.  c_2^{2} \left(2\text{Re}[\gamma_1] +\sqrt{3}\epsilon  (2\text{Im}[\gamma_1] +5 i) 
   -5\right)\right]+
    \nonumber\\  
&-6 X T  \left[c_1\left(2\text{Re}[\gamma_1] +\sqrt{3} \epsilon (2\text{Im}[\gamma_1] - 3i)  
   +1\right)+c_2\left(2\text{Re}[\gamma_1] -\sqrt{3} \epsilon(2\text{Im}[\gamma_1] +3i) 
   -5\right)\right]  +
    \nonumber\\  
&   
   +2X \left[2\text{Re}[\gamma_1]  \left(\sqrt{3} \epsilon -3i\right)-5\sqrt{3} \epsilon
   -3i\right]+
     \nonumber\\  
&     
+2 T \left[c_1\left(\text{Re}[\gamma_1]  \left(-\sqrt{3} \epsilon +3 i \right)-3\text{Im}[\gamma_1]  \left(3- i \sqrt{3}
   \epsilon \right)-2\sqrt{3} \epsilon +6i\right)+\right. \nonumber\\  
&+\left. c_2\left(\text{Re}[\gamma_1]  \left(-\sqrt{3} \epsilon +3 i \right)+3\text{Im}[\gamma_1]  \left(3- i \sqrt{3}
   \epsilon \right)+7 \sqrt{3} \epsilon -3
   i\right)\right]+
     \nonumber\\  
&   
   +3|\gamma_1|^2 -4\text{Re}[\gamma_1] -5i \sqrt{3} \epsilon\text{Re}[\gamma_1] +3\sqrt{3} \text{Im}[\gamma_1] +2i \sqrt{3} \epsilon -2   
\end{align}}
{\small \begin{align}
\label{M4}
&M_4 =12 X^4+12 T^4  \left(c_1^2-c_1c_2+c_2^2\right)^2 +36 T^2 X^2\left(c_1^2+c_2^2\right) -24 X^3 T (c_1+c_2) -24 X T^3\left(c_1^3+c_2^3\right)  +
\nonumber\\
& +8 \sqrt{3}\epsilon X^3-4 \sqrt{3} \epsilon T^3\left(4 c_1^3-3 c_1^2 c_2-3 c_1 c_2^2+4 c_2^3\right) + \nonumber\\ 
&-12 \sqrt{3}   \epsilon (c_1+c_2)X^2 T+12 \sqrt{3} \epsilon  X T^2 \left(3 c_1^2-4 c_1 c_2+3 c_2^2\right) +
 \nonumber\\ 
&+6 X^2(2\text{Re}[\gamma_1] +1) -6 T^2 \left[c_1^2 \left(\text{Re}[\gamma_1] -\sqrt{3} \epsilon\text{Im}[\gamma_1]  -7\right)+ 2 c_1 c_2 (5-2 \text{Re}[\gamma_1] )+\right. \nonumber\\ 
&+\left. c_2^2 \left(\text{Re}[\gamma_1]+\sqrt{3} \epsilon \text{Im}[\gamma_1]    -4\right)\right]  +
 \nonumber\\ 
&-12XT \left[c_1 \left(\text{Re}[\gamma_1] +\sqrt{3} \epsilon  \text{Im}[\gamma_1]  +2\right)+c_2 \left(\text{Re}[\gamma_1] -\sqrt{3}\epsilon
   \text{Im}[\gamma_1]   -1\right)\right]+2 \sqrt{3} \epsilon X (2 \text{Re}[\gamma_1] +1)  + \nonumber\\ 
&-2 T \left[c_1 \left(9 \text{Im}[\gamma_1] +\sqrt{3}\epsilon
   (\text{Re}[\gamma_1] +5)  \right)+c_2 \left(\sqrt{3} \epsilon (\text{Re}[\gamma_1] -4) -9
   \text{Im}[\gamma_1] \right)\right] +
 \nonumber\\      
&+3|\gamma_1|^2+3 \sqrt{3} \epsilon \text{Im}[\gamma_1]    -\text{Re}[\gamma_1] + 4  
\end{align} }
\end{subequations}


\begin{thebibliography}{99}

\bibitem{D2009}Degasperis A,  \textit{Multiscale expansion and integrability of dispersive wave equations}. In: Mikhailov A, \textit{Integrability}, vol. 767, pp. 215-244, Berlin, Springer (2009)

\bibitem{C1989} Calogero F, \textit{Universality and integrability of the nonlinear
PDEs describing N wave interactions}, J. Math. Phys. \textbf{30}, 28-40 (1989)

\bibitem{C1991} Calogero F, \textit{Why are certain nonlinear PDEs both widely applicable and integrable?} In 
\textit{What is integrability?} Zakharov V.E. (ed), Springer Verlag, 
Berlin, 1--62 (1991)

\bibitem{K76}
 Kaup D J, \textit{The three-wave interaction--a nondispersive phenomenon},
Stud. Appl. Math. \textbf{55}, 9 (1976).

\bibitem {D2011} Degasperis A, \textit{Integrable nonlocal wave interaction models},
 J. Phys. A: Math. Theor. \textbf{44}, 052002 (2011)
 
 \bibitem {BDDR2012} Borgna J P, Degasperis A, De Leo M F, Rial D, \textit{Integrability of nonlinear wave equations and solvability of their initial value problem},
  J. Math. Phys. \textbf{53}, 043701 (2012)
  
 \bibitem {AS1978} Ablowitz M J, Satsuma J, \textit{Solitons and rational solutions of nonlinear evolution equations},
  J. Math. Phys. \textbf{19}, 2180 (1978)
  
\bibitem {ASA2010} Ankiewicz A, Soto-Crespo J M, Akhmediev N, \textit{Rogue waves and rational solutions of the Hirota equation},
Phys. Rev. E  \textbf{81}, 046602 (2010)

 \bibitem{P1998} Pelinovsky D, \textit{Rational solutions of the KP hierarchy and the dynamics of their poles II},
J. Math. Phys. \textbf{39}, 5377-5395 (1998) 
  
 \bibitem {ACA2010} Ankiewicz A, Clarkson P A, Akhmediev N, \textit{Rogue waves, the pattern of their zeros and integral relations},
J. Phys. A: Math. Theor.  \textbf{43}, 122002 (2010)

\bibitem {DM2011} Dubard P, Matveev V B, \textit{Multi-rogue waves solutions to the focusing NLS equation and the KP-I equation}, 
Nat. Hazards Earth Syst. Sci., \textbf{11}, 667-672 (2011)

\bibitem {GLL2011} Guo B, Ling L, Liu Q P, \textit{Nonlinear Schr\"{o}dinger Equation: Generalized Darboux Transformation and Rogue Wave Solutions}
arXiv:1108.2867v2 (2011)
  
\bibitem {G2013} Gaillard P, \textit{Degenerate determinant representation of solutions of the nonlinear Schr\"{o}dinger equation, higher order Peregrine breathers and multi-rogue waves},
J. Math. Phys. \textbf{54}, 013504 (2013)

\bibitem{GL2011}Guo B, Ling L, \textit{Rogue Wave, Breathers and Bright-Dark-Rogue Solutions for the Coupled Schr\"{o}dinger Equations}
Chin. Phys. Lett. \textbf{28}, 110202 (2011)

\bibitem{BDCW2012} Baronio F, Degasperis A, Conforti M, Wabnitz S,
 \textit{Solutions of the Vector Nonlinear Schr\"{o}dinger Equations: Evidence
for Deterministic Rogue Waves},
 Phys. Rev. Lett. \textbf{109}, 044102-044106 (2012)
 
\bibitem{CS2013} Chen S, Song L, \textit{Rogue Waves in coupled Hirota systems},
Phys. Rev. E \textbf{87}, 032910 (2013) 

\bibitem{ZL2013} Zhao L, Liu J,  \textit{Rogue-Wave solutions of a three-component coupled nonlinear Schr\"{o}dinger equation},
Phys. Rev. E \textbf{87}, 013201 (2013)

\bibitem{P1983} Peregrine D H,  \textit{Water waves, nonlinear Schr\"{o}dinger equations and their solutions},
J. Aust. Math. Soc. B \textbf{25}, 16-43 (1983)

\bibitem{KPS2009} Kharif C, Pelinovsky E, Slunyaev A, \textit{Rogue Waves in the Ocean}, Springer-Verlag (2009)

\bibitem{CHA2011} Chabchoub A, Hoffmann N P, Akhmediev N, \textit{ }, \textit{Rogue Wave Observation in a Water Wave Tank}
Phys. Rev. Lett. \textbf{106}, 204502 (2011)

\bibitem{KFFMDGAD2010} Kibler B, Fatome J, Finot C, Millot G, Dias F, Genty G,  Akhmediev N, Dudley J M, Nat. Phys. \textbf{6}, 790 (2010)

\bibitem{BSN2011} Bailung H, Sharma S K, Nakamura Y, \textit{Observation of Peregrine Solitons in a Multicomponent Plasma with Negative Ions}
Phys. Rev. Lett. \textbf{107}, 255005 (2011)

\bibitem{SS2009} Stenflo L, Shukla P K, \textit{Nonlinear acoustic-gravity waves}, 
J. Plasma Phys. \textbf{75}, 841 (2009)

\bibitem{GEKMM2008} Ganshin A N, Efimov V B, Kolmakov G V, Mezhov-Deglin L P, McClintock P V E,  \textit{Observation of an Inverse Energy Cascade in Developed Acoustic Turbulence in Superfluid Helium}
Phys. Rev. Lett. \textbf{101}, 065303 (2008)

\bibitem{BKA2009} Bludov Y V, Konotop V V,  Akhmediev N, 
\textit{Matter rogue waves},
Phys. Rev. A \textbf{80}, 033610 (2009)

\bibitem{SPX2010} Shats M, Punzmann H, Xia H, 
\textit{Capillary Rogue Waves},
Phys. Rev. Lett. \textbf{104}, 104503 (2010)
 
\bibitem{DL2007} Degasperis A and Lombardo S, \textit{Multicomponent integrable wave equations I. Darboux--Dressing Transformation}, J. Phys. A: Math. Theor. \textbf{40}, 961--977 (2007)

\bibitem{DL2009} Degasperis A and Lombardo S, \textit{Multicomponent integrable wave equations II. Soliton solutions}, J. Phys. A: Math. Theor. \textbf{42}, 385206 (2009)

\end{thebibliography}
\end{document}